\pdfoutput=1
\documentclass[twocolumn]{article}

\usepackage[modulo,switch]{lineno}
\modulolinenumbers[1]

\usepackage[utf8]{inputenc}
\usepackage[T1]{fontenc}
\usepackage{hyperref}
\usepackage{amsmath}
\usepackage{amssymb}
\usepackage{graphicx}
\usepackage{subcaption}
\usepackage{xcolor}
\usepackage{stmaryrd}
\usepackage{empheq}

\makeatletter\@ifundefined{date}{}{\date{}}
\makeatother

\evensidemargin\oddsidemargin \hoffset-22.4mm \voffset-28.4mm
\newcommand{\nin}{\noindent}

\newcommand{\pr}[1]{\left( #1 \right)}
\newcommand{\hk}[1]{\left[ #1 \right]}
\newcommand{\dhk}[1]{\llbracket #1 \rrbracket}
\newcommand{\br}[1]{\left\{ #1 \right\}}
\newcommand{\md}[1]{\left| #1 \right|}

\newcommand{\gu}[1]{``#1''}

\newcommand{\PN}{\text{PN}}
\newcommand{\GN}{\text{GN}}
\newcommand{\RD}{\text{R2}}
\newcommand{\R}{\mathbb{R}}
\newcommand{\C}{\mathbb{C}}

\newcommand{\w}{\omega}

\newcommand{\Ze}{Z_{\text{exp}}}
\newcommand{\Zr}{Z_{\text{fit}}}
\newcommand{\wl}{\omega_\ell}

\newcommand{\fl}{f_\ell}
\newcommand{\Ql}{Q_\ell}
\newcommand{\gaml}{\frac{\wl}{\Ql}}

\newcommand{\ps}[1]{p_{\text{thresh}}^{\text{#1}}}
\newcommand{\fs}[1]{f_{\text{thresh}}^{\text{#1}}}

\newcommand{\pso}[1]{p_{\text{thresh},n}^{\text{opt,#1}}}
\newcommand{\fso}[1]{f_{\text{thresh},n}^{\text{opt,#1}}}
\newcommand{\flo}[1]{f_{\ell,n}^{\text{opt,#1}}}

\newcommand{\flon}[2]{f_{\ell,#2}^{\text{opt,#1}}}

\newcommand{\wlql}{\frac{\wl}{\Ql}}

\newcommand{\X}{\mathbf{X}}

\renewcommand{\Re}{\mathfrak{Re}}
\renewcommand{\Im}{\mathfrak{Im}}

\newcommand{\J}{\mathrm{j}}

\DeclareMathOperator{\sgn}{\text{sgn}}

\newcommand{\der}[2]{\frac{\mathrm{d}#1}{\mathrm{d}#2}}
\newcommand{\dder}[2]{\frac{\mathrm{d}^2#1}{\mathrm{d}#2^2}}

\begin{document}

\title{Diversity of ghost notes in tubas, euphoniums and saxhorns}

\author{Rémi Mattéoli$^{*}$, Joël Gilbert$^{*}$, Soizic Terrien$^{*}$,\\ Jean-Pierre Dalmont$^{*}$, Christophe Vergez$^{\dagger}$, Sylvain Maugeais$^{\ddagger}$, Emmanuel Brasseur$^{*}$ \\
$^{*}$ Laboratoire d'Acoustique de l'Université du Mans (LAUM), UMR CNRS 6613,\\ Institut d'Acoustique - Graduate School (IA-GS), CNRS, Le Mans Université, France\\
$^{\dagger}$ Aix Marseille Univ, CNRS, Centrale Marseille, LMA, UMR 7031, France\\
$^{\ddagger}$ Laboratoire Manceau de Mathématiques – Le Mans Université, 72085 Le Mans, France}

\maketitle\thispagestyle{empty}

\begin{abstract}
    The ghost note is a natural note which can be played exclusively on bass brass instruments with a predominantly-expanding bore profile such as tubas, euphoniums or saxhorns. It stands between the pedal note -- the lowest natural note playable, or first regime -- and the instrument's second regime. However, if the interval between the pedal note and the second regime remains close to an octave regardless of the instrument, the interval between the pedal note and the ghost note vary from a minor third to a perfect fourth. References about this note are very scarce, and it is not commonly known among tuba players.

    This study shows that an elementary brass model describing the player coupled to the instrument is capable of bringing both the ghost and the pedal note to light. Here, we adopt a dynamical systems point of view and perform a bifurcation analysis using a software of numerical continuation. The numerical results provided in terms of frequency intervals between pedal note and ghost note are compared with frequency intervals experimentally inferred from recordings of seven different types of tuba, each of them being played by two professional tuba players.
\end{abstract}

\section{Introduction}

One important goal of the acoustics of wind instruments is to describe and quantify the intonation and ease of playing of an instrument. From the physics point of view, a classical approach consists in modelling the coupled system composed of the musician and the instrument. Of particular interest is the influence of the musician's control parameters on the oscillation frequency (linked to the intonation), and the minimum blowing mouth pressure required to trigger self-oscillations (related to the ease of playing). Indeed, it is assumed here that the musician's feeling of ease of playing partly relies on the blowing threshold pressure: the higher the latter, the higher the physical effort the player has to make to play a note \cite{campbellScienceBrassInstruments2021}. In practice, the musician can play several notes without changing the acoustical properties of the instrument itself, that is to say without depressing any valves in the case of a tuba, or moving the slide in the case of a trombone. These notes are called the natural notes (B$\flat$1, B$\flat$2, F3, B$\flat$3, D4, F4,... in the case of a trombone or a B$\flat$-bass saxhorn for instance), and their frequencies are close to the resonance frequencies of the instrument as a whole, except for the lowest note playable (B$\flat$1, $\text{frequency} \approx 58\,\text{Hz}$).

The oscillation regimes and ease of playing of bass brass instruments have been investigated in \cite{matteoliMinimalBlowingPressure2021}: a bifurcation analysis showed that a simple mathematical model accurately describes all the natural notes of a bass trombone and a euphonium in terms of frequency. However, a main difference was highlighted in \cite{velutHowWellCan2017,matteoliMinimalBlowingPressure2021}, between instruments with predominantly-cylindrical bore profile such as the trombone, and instruments with predominantly-expanding bore profile such as the euphonium or bass saxhorn. Indeed, the latter exhibit an extra regime located between the pedal note and the second regime, referred to as \gu{ghost note}. It was also pointed out that the frequency interval between the pedal note and the ghost note could vary between a minor third and a perfect fourth, depending on the bore geometry of the instrument, whereas all other intervals between natural notes remained approximately the same.

\begin{figure}[h!]
    \centering
    \reflectbox{\includegraphics[width=\linewidth]{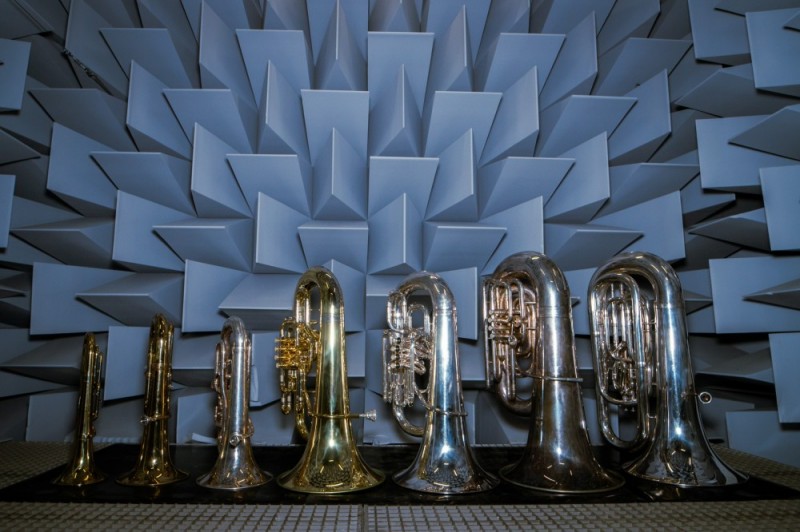}}
    \caption{Photograph showing the different tubas studied in the present publication. From left to right (lowest to highest nominal pitch): contrabass tuba in B$\flat$, contrabass tuba in C, bass tuba in E$\flat$, bass tuba in F, euphonium in B$\flat$, bass saxhorn in B$\flat$, baritone saxhorn in B$\flat$. Details about each tuba can be found in the table \ref{carac_tubas}.}
    \label{all_tubas}
\end{figure}

This paper aims at proving that the same generic brass model is also able to render the variety of ghost notes existing among the tuba family. In that respect, a bifurcation analysis is performed to determine the easiest-to-play pedal note, ghost note and second regime. Then, the frequency intervals between the pedal note and the second regime, and between the ghost note and the second regime are assessed for the seven tubas displayed in figure \ref{all_tubas}, and compared to recordings of two professional tuba players playing the same set of tubas. A more exhaustive description of the characteristics of each tuba and their mouthpiece can be found in table \ref{carac_tubas}. It is worth noting from this table that the baritone saxhorn, the bass saxhorn and the euphonium have the same nominal pitch (B$\flat$), and the same set of natural notes. However, while the bass saxhorn and the euphonium have a bore profile quite similar to the other tubas (predominantly-expanding bore profile), the baritone saxhorn's bore profile is intermediate between a tuba's and a trombone's bore profile (predominantly-cylindrical bore profile), as it is seen in the bottom plot of figure \ref{comp_tb_sb_eu}. Therefore, this instrument constitutes a borderline case in this paper.

\begin{table*}[h!]
    \centering
    \begin{tiny}
        \begin{tabular}{|c|c|c|c|c|c|c|c|}
            \hline
            & B$\flat$-contrabass tuba & C-contrabass tuba & E$\flat$-bass tuba & F-bass tuba & B$\flat$-euphonium & B$\flat$-bass saxhorn & B$\flat$-baritone saxhorn\\
            \hline
            Nominal pitch & 18-ft B$\flat$ & 16-ft C & 13-ft E$\flat$ & 12-ft F & 9-ft B$\flat$ & 9-ft B$\flat$ & 9-ft B$\flat$\\
            \hline
            Brand  & Besson & Hirsbrunner & Besson & Melton & Besson & A. Courtois & SML\\
            \hline
            Model & BE994 & HBS 510 & BE983 & 2250 & BE967 & AC164 & BA16\\
            \hline
            Mouthpiece & Perantucci PT-50 & Perantucci PT-50 & Romera FT20 & Romera FT20 & A. Courtois T2 & A. Courtois T2 & A. Courtois T2\\
            \hline
            Visual & \includegraphics[height=3cm]{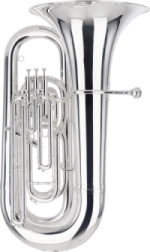} & \includegraphics[height=3cm]{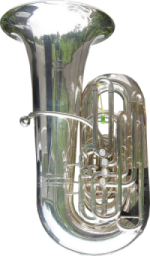} & \includegraphics[height=2.786cm]{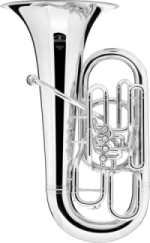} & \includegraphics[height=2.829cm]{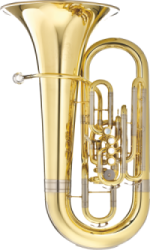} & \includegraphics[height=2.226cm]{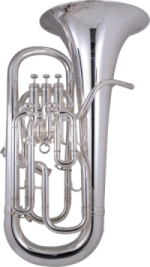} & \includegraphics[height=2.226cm]{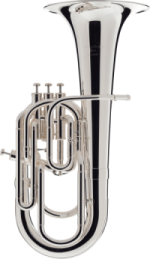} & \includegraphics[height=1.971cm]{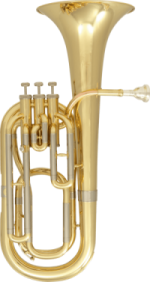}\\
            \hline
        \end{tabular}
    \end{tiny}
    \caption{Characteristics of the seven tubas studied in this paper, classified in the same order as in figure \ref{all_tubas}. By convention, the nominal pitch specifies a number of feet (ft) coupled with a musical pitch name, being the nominal fundamental of the instrument (lowest note playable without any valve depressed). The number of feet corresponds to the equivalent cone length $L^{\text{ec}}_{\text{nom}}$, related to the frequency $F_{\text{nom}}$ of its nominal fundamental by the formula $L^{\text{ec}}_{\text{nom}} = \frac{c_0}{2F_{\text{nom}}}$, with $c_0$ the speed of sound in the air\protect\footnotemark.}
    \label{carac_tubas}
\end{table*}

\footnotetext{See for instance paragraph 1.2.3 of \cite{campbellScienceBrassInstruments2021} for more details about the notion of nominal pitch.}

In section \ref{TB}, we describe the considered model and numerical methods, namely linear stability analysis and numerical continuation. In section \ref{MBPPF}, we use linear stability analysis and bifurcation diagrams, respectively, to investigate the minimal mouth pressure and playing frequencies of two particular tubas. Eventually, we present in section \ref{CBTWDTL} the procedure used to assess experimental oscillation frequencies based on recordings of two professional tuba players, which we use for comparison with numerical oscillation frequencies.

\section{Theoretical background}
\label{TB}

\subsection{Physical model of brass instrument}
\label{GBM}

This subsection briefly presents the brass instrument model considered throughout the article. It has already been thoroughly described in \cite{mcintyreOscillationsMusicalInstruments1983,hirschbergMechanicsMusicalInstruments1995,velutHowWellCan2017,matteoliMinimalBlowingPressure2021}. Brass instruments as a whole can be described through both linear and nonlinear mechanisms. More precisely, a localised nonlinear element (the lips' valve effect, namely the velocity modulation caused by the lips' vibration) excites a passive linear acoustic multimode element (the musical instrument, usually characterised by its input impedance in the frequency domain) \cite{fletcherPhysicsMusicalInstruments1998}. The latter acoustic resonator exerts, in turn, a retroaction on the former mechanical resonator. Such musical instruments are self-sustained oscillators: they generate an oscillating acoustic pressure (the note played) from a static overpressure in the player's mouth (the blowing pressure) \cite{chaigneAcousticsMusicalInstruments2016,campbellScienceBrassInstruments2021,fletcherPhysicsMusicalInstruments1998}.

\begin{figure}[h!]
    \centering
    \includegraphics[width=0.6\linewidth]{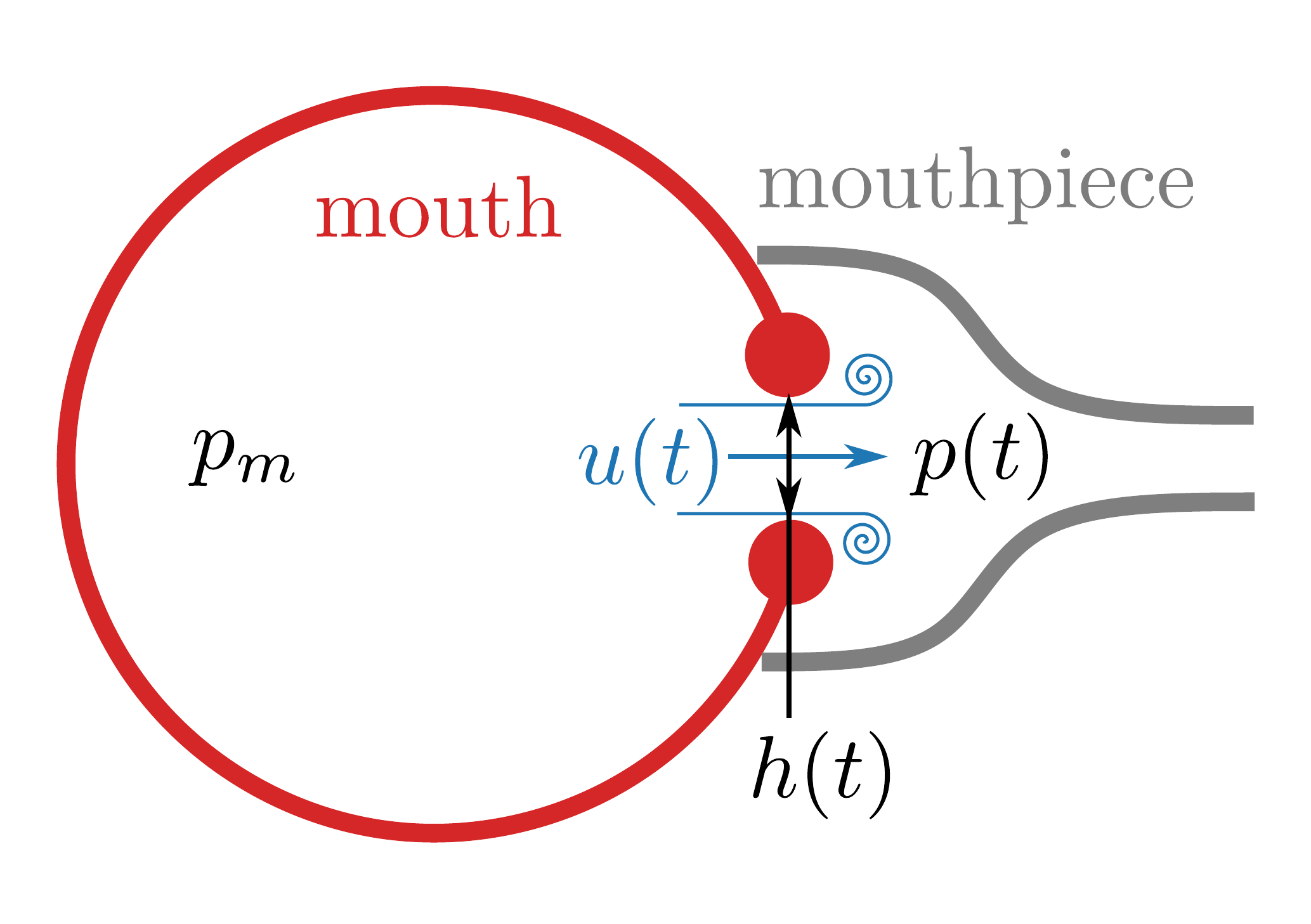}
    \caption{Schematic representation of the player's mouth coupled to the mouthpiece of a brass instrument. $p_m$ stands for the blowing mouth pressure, $u(t)$ is the volume velocity flowing from the player's mouth into the mouthpiece, $p(t)$ is the acoustic pressure inside the mouthpiece, and $h(t)$ is the height between the lips.}
    \label{schema_mouthpiece}
\end{figure}

The brass instrument coupled to the player is then described by a system of three equations. Because it relies on major simplifications \cite{campbellScienceBrassInstruments2021,elliottRegenerationBrassWind1982}, this model is often referred to as \gu{elementary}\footnote{The vibrating lips are modeled as a linear one-degree-of-freedom oscillator. The upstream resonances of the player’s windway are neglected, as is nonlinear propagation of sound in the air column of the instrument. Wall vibrations are also ignored.}. More precisely, the three equations link the lip-opening height $h(t)$, the pressure in the mouthpiece $p(t)$ and the volume velocity entering the instrument $u(t)$, which are the three independent variables of interest in this paper. These are all represented schematically in figure \ref{schema_mouthpiece}. Several control parameters (that is to say controlled by the musician) are involved: this includes, in particular, the blowing mouth pressure $p_m$ -- which is simply referred to as the mouth pressure in the following -- and the lips' resonance frequency $\fl$.


First, the vibrating lips of the musician are described by a one-degree-of-freedom damped oscillator \cite{fletcherAutonomousVibrationSimple1993}:

\begin{equation}
    \dder{h}{t} + \wlql \der{h}{t} + \wl^2 \pr{h(t)-H(\wl)} = \frac{p_m-p(t)}{\mu},\\
    \label{eqh}
\end{equation}

\nin
where $\wl = 2\pi\fl$ and $\Ql$ are the angular resonance frequency and the quality factor of the lips, respectively, $\mu$ is the lips' mass per unit area, and $H(\wl)$ is the lip-opening height at rest. Based on \textit{in vivo} measurements of $H$ with respect to $\wl$ from \cite{elliottRegenerationBrassWind1982}, $H$ is assumed to be proportional to the inverse of the lips' resonance frequency as in \cite{matteoliMinimalBlowingPressure2021}. In this respect, we write $H(\wl) = 2\pi k/\wl$, $k$ being given in table \ref{lips_params} along with the other parameters of the model.\\

The Bernoulli's principle is applied between the mouth and the mouthpiece. Taking into account the pressure drop caused by the presence of turbulence in the mouthpiece, one obtains the following equation:

\begin{equation}
    u(t) = wh^+(t) \sgn\pr{p_m-p(t)} \sqrt{\frac{2\md{p_m-p(t)}}{\varrho}},\\
    \label{equ}
\end{equation}

\nin
with $w$ the lip-opening width (considered to be constant) and $\varrho$ the air density. Here, $h^+ = \max(h,0)$ accounts for the fact that the lips cannot physically interpenetrate: as soon as the lips touch ($h = 0$), the volume velocity is forced to zero. The sign function $\sgn$ accounts for the possibility of air flowing from the instrument into the player's mouth.\\

Finally, the acoustic input impedance $Z(\w)$ of the resonator is described in the Fourier domain as the ratio between the acoustic pressure and the volume velocity at the input of the instrument. This provides another link between the mouth pressure and the volume velocity:

\begin{equation}
    P(\w) = Z(\w) U(\w),
    \label{eqZ}
\end{equation}

\nin
with $\w$ the angular frequency. Figure \ref{zc_BE994} shows the modulus and phase of $Z$ with respect to the frequency $f$, for the B$\flat$-contrabass tuba (see table \ref{carac_tubas} for details) which will be considered in subsection \ref{CCT}.

%
\begin{table}[h!]
    \centering
    \begin{tabular}{|c|c|c|}
        \hline
        $w\,[\text{m}]$ & $\Ql$ & $\frac{1}{\mu}\,[\text{m}^2\,\text{kg}^{-1}]$\\
        \hline
        $1.2\times10^{-2}$ & $7.0$ & $1.1\times10^{-1}$\\
        \hline
        \hline
        $\varrho\,[\text{kg}\,\text{m}^{-3}]$ & $c_0\,[\text{m}\,\text{s}^{-1}]$ & $k\,[\text{m}\,\text{Hz}]$\\
        \hline
        $1.2$ & $3.4\times10^2$ & $7.4\times10^{-2}$\\
        \hline
    \end{tabular}
    \caption{Lips' parameters used throughout the article, taken from \cite{matteoliMinimalBlowingPressure2021}.}
    \label{lips_params}
\end{table}

\subsection{Numerical considerations}
\label{NT}

We aim here at determining the minimal mouth pressure for which each periodic regime (natural notes B$\flat$1, B$\flat$2, F3, B$\flat$3, D4, F4,... in the case of an euphonium) is observable, as well as the corresponding playing frequencies.

In this subsection, the numerical methods allowing for the investigation of the periodic solutions of model $\br{(\ref{eqh}) \cup (\ref{equ}) \cup (\ref{eqZ})}$ are presented, and the modal representation \cite{pagneuxStudyWavePropagation1996} of the input impedance required for the practical implementation of these methods is discussed.

\subsubsection{Numerical methods}
\label{NM}

Several numerical methods are available to investigate the influence of a control parameter -- such as $p_m$ or $\fl$, which are the main control parameters of the musician -- on the dynamics of the system. Linear stability analysis \cite{velutHowWellCan2017,seydelPracticalBifurcationStability2010,fletcherAutonomousVibrationSimple1993} consists in studying the stability of the system linearised around an equilibrium solution, obtained by zeroing all time derivatives. This allows one to determine the threshold value of the control parameter at which the equilibrium solution destabilises. This happens here in a Hopf bifurcation \cite{kuznetsovElementsAppliedBifurcation2004}, which means that two complex conjugate eigenvalues of the linearised system cross the imaginary axis. At this point, a small-amplitude periodic regime (stable or unstable) can emerge. This method has already been applied to physical models of musical instruments, such as woodwind instruments in \cite{wilsonOperatingModesClarinet1974,changReedStability1994,silvaInteractionReedAcoustic2008,karkarOscillationThresholdClarinet2012,saneyoshiFeedbackOscillationsReed1987}, flute-like instruments in \cite{terrienWhatExtentCan2014}, and brass instruments in \cite{cullenBrassInstrumentsLinear2000a,lopezPhysicalModelingBuzzing2006,silvaOscillationThresholdsStriking2007a}

In the case of brass instruments, both $p_m$ and $\fl$ are considered as control parameters. Figure \ref{asl_BE994} (blue curve) represents the pressure threshold predicted by the linear stability analysis with respect to $\fl$. This analysis also provides the oscillation frequency at threshold (referred to as the threshold frequency in the following), that is to say the frequency of the periodic regime that emerges when the equilibrium solution destabilises. This method is applied in subsection \ref{LSA} to the case of the B$\flat$-contrabass tuba.

One of the main advantages of this method is its straight-forward implementation. However, it gives very little information on the oscillation regime of the model far from the so-called Hopf bifurcation point at which the equilibrium loses its stability. To know more about the oscillating solution arbitrarily far from the bifurcation point (namely the point at which the equilibrium solution becomes unstable), another approach -- not considered in this article -- consists in numerically solving the whole system, thanks to an ODE solver \cite{velutHowWellCan2017,gilbertMinimalBlowingPressure2020,silvaMoReeSCFrameworkSimulation2014}. In doing so, both the transient and the stationary parts of the solution are obtained for any value of the control parameter. Nevertheless, this approach becomes tedious and unsuitable in the context of the systematic investigation of the influence of the control parameter on the oscillation regime. It is also unsure that all regimes are described with this method, due to sensitivity to initial conditions.

Continuation methods, on the other hand, are more suitable to gain access to an extensive knowledge of all the oscillating regimes of the system, that is to say a family of periodic solutions with respect to a parameter of interest for instance. It consists in computing the waveform of the oscillating solution for successive values of the control parameter. The waveform corresponding to a new value of the control parameter is then deduced from previously computed waveforms, through a predictor/corrector algorithm. The behaviour of an oscillating solution of the system with respect to a control parameter such as $p_m$ is then assessed by plotting bifurcation diagrams, which are shown in section \ref{MBPPF}. This approach is implemented in several softwares dedicated to advanced numerical bifurcation analysis, such as AUTO \cite{doedelAUTO97Continuation1999}, which is used in this publication and in \cite{gilbertMinimalBlowingPressure2020,akayContinuationAnalysisNonlinear2021} as well. While AUTO relies on the collocation method, other softwares -- such as Manlab \cite{guillotTaylorSeriesbasedContinuation2019,cochelinHighOrderPurely2009,karkarHighorderPurelyFrequency2013,colinotInfluenceGhostReed2019a,freourNumericalContinuationPhysical2020} -- are based on the Harmonic Balance Method \cite{nakhlaPiecewiseHarmonicBalance1976}. It is worth noting that continuation methods require the system to be written in the form $\der{\X}{t} = F(\X)$, with certain smoothness properties on $F$. Therefore, some work has yet to be done on the equations presented in section \ref{GBM}, which is done in the following.

\subsubsection{Input impedance}

\begin{figure}[h!]
    \centering
    \subcaptionbox{Input impedance as a function of frequency. \label{zc_BE994}}{\includegraphics[width=0.55\linewidth]{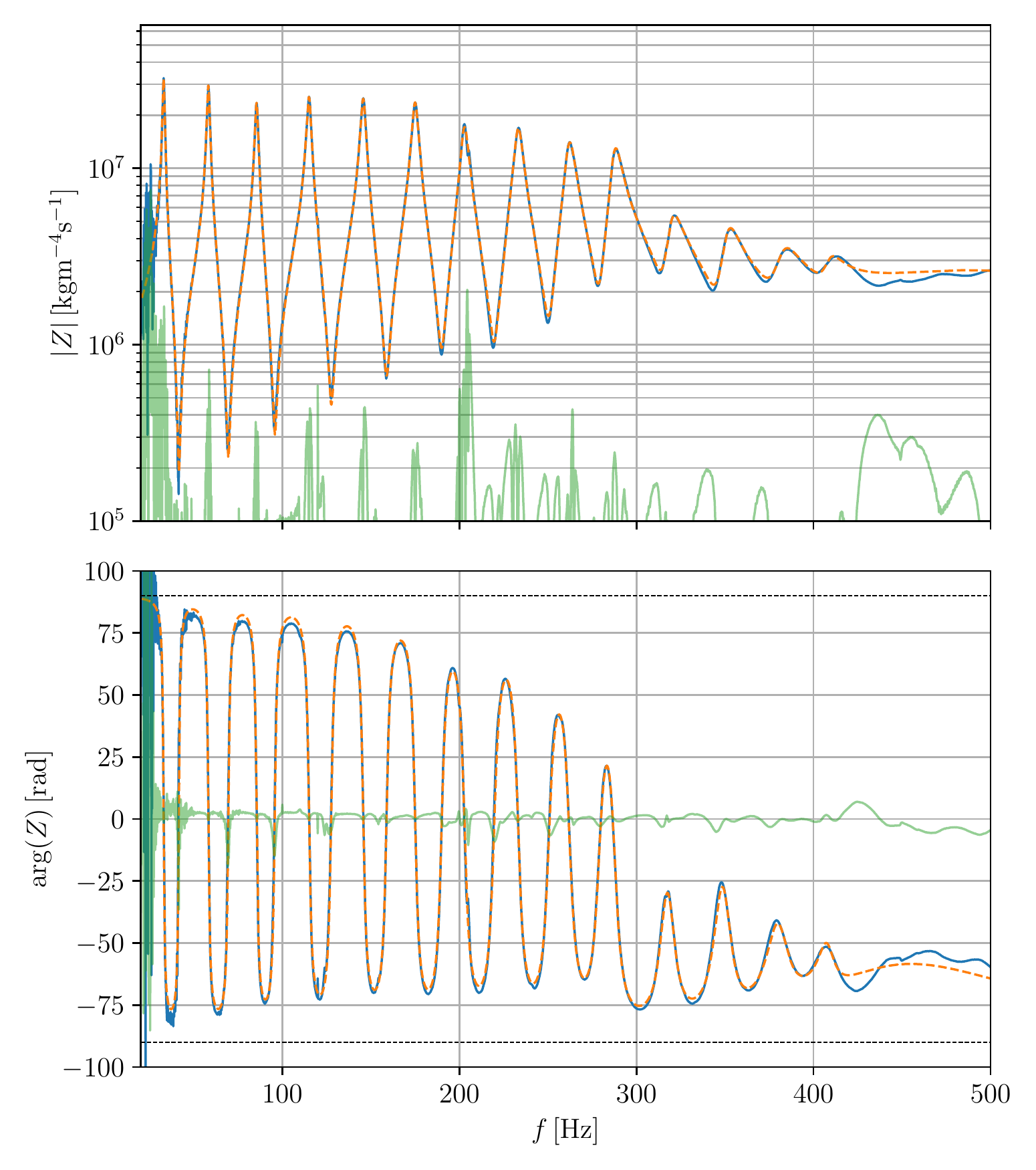}}
    \subcaptionbox{Experimental set-up. \label{z_bridge}}{\includegraphics[width=0.4\linewidth]{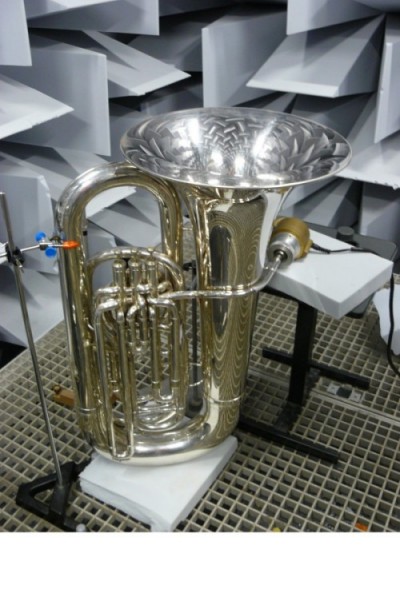}}
    \caption{Left: modulus (top) and phase (bottom) of the input impedance of the B$\flat$-contrabass tuba with respect to frequency. Blue curve: measured impedance; dashed orange curve: modal fit function with $N=13$ modes; green curve: error between measured and fitted impedances, defined by $\md{\md{\Zr}-\md{\Ze}}$ and $\arg\pr{\Zr}-\arg\pr{\Ze}$ as regards the modulus and phase, respectively. Right: photograph of the experimental set-up for the measure of the input impedance of the B$\flat$-contrabass tuba, displaying the device described in \cite{macalusoTrumpetNearperfectHarmonicity2011}, here mounted on the tuba's mouthpiece.}
    \label{mesimp_BE994}
\end{figure}

The implementation of the continuation method requires to rewrite equation (\ref{eqZ}) in the time domain. An analytical form of the input impedance $Z$ is therefore required to perform an inverse Fourier transform of (\ref{eqZ}). This quantity is quite easily measured, and is represented in figure \ref{zc_BE994} (blue curve). This measured impedance is then numerically fitted, in the frequency domain, by a sum of $N$ individual acoustical resonance modes of the following form \cite{ablitzerPeakpickingIdentificationTechnique2021,silvaMoReeSCFrameworkSimulation2014}:

\begin{equation}
    \Zr(\w) = \sum_{n=1}^N \frac{\J\w A_n}{\w_n^2-\w^2 + 2\J\xi_n\w_n\w}.
    \label{ZR_FA}
\end{equation}

\noindent Here, $\pr{A_n, \w_n, \xi_n} \in \R^3$ are the modal parameters of the modal decomposition. For numerical reasons, this set of parameters has been converted here into another one $\pr{C_n, s_n}$ so that the fitted input impedance is written:

\begin{equation}
    \Zr(\w) = Z_c \sum_{n=1}^N \pr{\frac{C_n}{\J\w-s_n} + \frac{C_n^*}{\J\w-s_n^*}},
    \label{ZC}
\end{equation}

\noindent where $z^*$ stands for the complex conjugate of $z$, $Z_c$ is the characteristic input impedance of the resonator defined as $Z_c = \frac{\varrho c_0}{S_e}$ with $S_e$ the input cross-sectional area, and $\pr{C_n,s_n} \in \C^2$ defined as follows:

\begin{equation}
    \left\{
        \begin{array}{r c l}
            C_n &=& \frac{A_n}{2}\pr{1 + \J\frac{\xi_n}{\sqrt{1-\xi_n^2}}},\\
            s_n &=& \w_n\pr{-\xi_n + \J\sqrt{1-\xi_n^2}}.\\
        \end{array}
    \right.
    \label{rel_CR}
\end{equation}

The new coefficients $\pr{C_n, s_n}$ also verify the relation $\Re\pr{C_n s_n^*} = 0$, since one switched from a 3-real parameter description with $\pr{A_n,\w_n,\xi_n}$ to a 2-complex parameter description with $\pr{C_n,s_n}$. The fitted impedance $\Zr(\w)$ is plotted (orange dashed curve) in figure \ref{zc_BE994}. 

Reinjecting equation (\ref{ZC}) in equation (\ref{eqZ}) and applying an inverse Fourier transform\footnote{For details about this calculation, see appendix A of \cite{matteoliMinimalBlowingPressure2021}.} leads to the following expression in the time domain \cite{freourNumericalContinuationPhysical2020,silvaMoReeSCFrameworkSimulation2014}:

\begin{equation}
    \der{p_n}{t} = Z_c C_n u + s_n p_n, \; n \in \dhk{1,N}.
    \label{eqZtemp}
\end{equation}

The mouthpiece pressure $p$ is then written as $p = 2\sum_{n=1}^{N} \Re\pr{p_n}$. Eventually, the system $\br{(\ref{eqh}) \cup (\ref{equ}) \cup (\ref{eqZtemp})}$ is now written in the form $\der{\X}{t} = F(\X)$, with:

\begin{equation}
    \begin{split}
        \X &= \pr{\br{X_m}_{m \in \dhk{1,2(N+1)}}}\\
           &= \pr{ h; \der{h}{t}; \br{\Re(p_n)}_{n \in \dhk{1,N}}; \br{\Im(p_n)}_{n \in \dhk{1,N}}}
    \end{split}
\end{equation}

\nin
the state vector, so that $h = X_1$, $\der{h}{t} = X_2$, and $p = 2\sum_{n=1}^{N} \Re\pr{p_n} = 2\sum_{n=3}^{N+2}X_n$. Taking the real and imaginary part of the $N$ equations (\ref{eqZtemp}) yields $2N$ real equations, so that the nonlinear function $F$ is defined as:

\begin{equation}
    F:\X\mapsto
    \begin{pmatrix}
        X_2\\
        - \gaml X_2 - \wl^2\pr{X_1-H} + \frac{p_m-2\sum_{n=3}^{N+2}X_n}{\mu}\\
        \Re\hk{s_1\pr{X_3 + \J X_{N+3}} + Z_c C_1 u(\X)}\\
        \Re\hk{s_2\pr{X_4 + \J X_{N+4}} + Z_c C_2 u(\X)}\\
        \vdots\\
        \Re\hk{s_N\pr{X_{N+2} + \J X_{2(N+1)}} + Z_c C_N u(\X)}\\
        \Im\hk{s_1\pr{X_3 + \J X_{N+3}} + Z_c C_1 u(\X)}\\
        \Im\hk{s_2\pr{X_4 + \J X_{N+4}} + Z_c C_2 u(\X)}\\
        \vdots\\
        \Im\hk{s_N\pr{X_{N+2} + \J X_{2(N+1)}} + Z_c C_N u(\X)}\\
    \end{pmatrix},
    \label{sys}
\end{equation}

with $u:\X \mapsto w X_1^+ \sgn\pr{p_m-2\sum_{n=3}^{N+2}X_n} \times$ $\sqrt{\frac{2}{\varrho}\md{p_m-2\sum_{n=3}^{N+2}X_n}}$.

\subsubsection{Regularization of the volume velocity}

Continuation methods rely on the assumption of a smooth nonlinear vector function $F$, which needs to be at least $\mathcal{C}^1$. Consequently, equation (\ref{equ}) is regularised in the exact same way as in \cite{matteoliMinimalBlowingPressure2021,colinotInfluenceGhostReed2019a}:

\begin{equation}
    u \underset{\eta \to 0}{\sim} w \times \frac{h + h_0\sqrt{\pr{\frac{h}{h_0}}^2+\eta}}{2} \times \frac{p_m-p}{\sqrt{p_0}\sqrt[4]{\pr{\frac{p_m-p}{p_0}}^2+\eta}},
    \label{equreg}
\end{equation}

or equivalently

\begin{equation}
    \begin{split}
        u &\underset{\eta \to 0}{\sim} w \times \frac{X_1 + h_0\sqrt{\pr{\frac{X_1}{h_0}}^2+\eta}}{2}\\
          &\hspace{5mm}\times \frac{p_m-2\sum_{n=3}^{N+2}X_n}{\sqrt{p_0}\sqrt[4]{\pr{\frac{p_m-2\sum_{n=3}^{N+2}X_n}{p_0}}^2+\eta}}
    \end{split}
    \label{equregX}
\end{equation}

\noindent
in terms of the components of the state vector $\X$, with $\eta$ the regularization parameter, which is fixed to $10^{-4}$ in the following. $h_0 = 5\times10^{-4}\,\text{m}$ is defined for purely dimensional reasons, and $p_0$ is defined similarly to the closure pressure for woodwind instruments \cite{gilbertMinimalBlowingPressure2020}: $p_0 = \mu \w_0^2 h_0$, except that $\w_0$ is chosen close to the first resonance frequency of the resonator. Indeed, the choice of the lips' angular resonance frequency $\wl$ generally considered for woodwind instruments is not suitable in the case of brass instruments where $\wl$ is no longer constant. In practice, the choice is not $\w_0 = \w_{\text{res},1}$ (where $\br{\w_{\text{res},n}}_{n\in\dhk{1,N}}$ are the resonance angular frequencies of the resonator), but rather $\w_0 = \w_{\text{res,4}}/4$, because the fourth resonance of instruments having the same nominal pitch (a trombone and a euphonium for instance) appears to be quite constant, in contrast to the first resonance frequency which varies up to 8 semitones between a trombone and euphonium.

In the following, the system of equations (\ref{sys}) with $u(\X)$ given by equation (\ref{equregX}) are processed numerically in a dimensionless form, which is detailed in appendix B of \cite{matteoliMinimalBlowingPressure2021}. Typical bifurcation diagrams of the system will be shown in section \ref{MBPPF}.

\section{Minimal blowing pressures and playing frequencies}
\label{MBPPF}

It was shown in \cite{velutHowWellCan2017,matteoliMinimalBlowingPressure2021} that the linear stability analysis was able to accurately describe the ghost note in terms of playing frequency. However, this method did not allow to describe the pedal note of any tuba. Therefore, this section aims at proving that the pedal note can actually be described by the generic brass model analysed through a complete bifurcation analysis instead of a linear stability analysis only. This section focuses on the first three natural notes of the B$\flat$-contrabass tuba and the B$\flat$-baritone saxhorn, as they show a variety of different behaviours of the system.

\subsection{Case of the B$\flat$-contrabass tuba}
\label{CCT}

This subsection focuses on the study of the first three natural notes of the B$\flat$-contrabass tuba (B$\flat$0, D1, B$\flat$1, see figure \ref{partiels_BE994_gn}), which is the lowest-pitched tuba among the seven tubas, namely the one with the longest tube (see table \ref{carac_tubas}).

\subsubsection{Linear stability analysis}
\label{LSA}

\begin{figure}[h!]
    \centering
    \includegraphics[width=\linewidth]{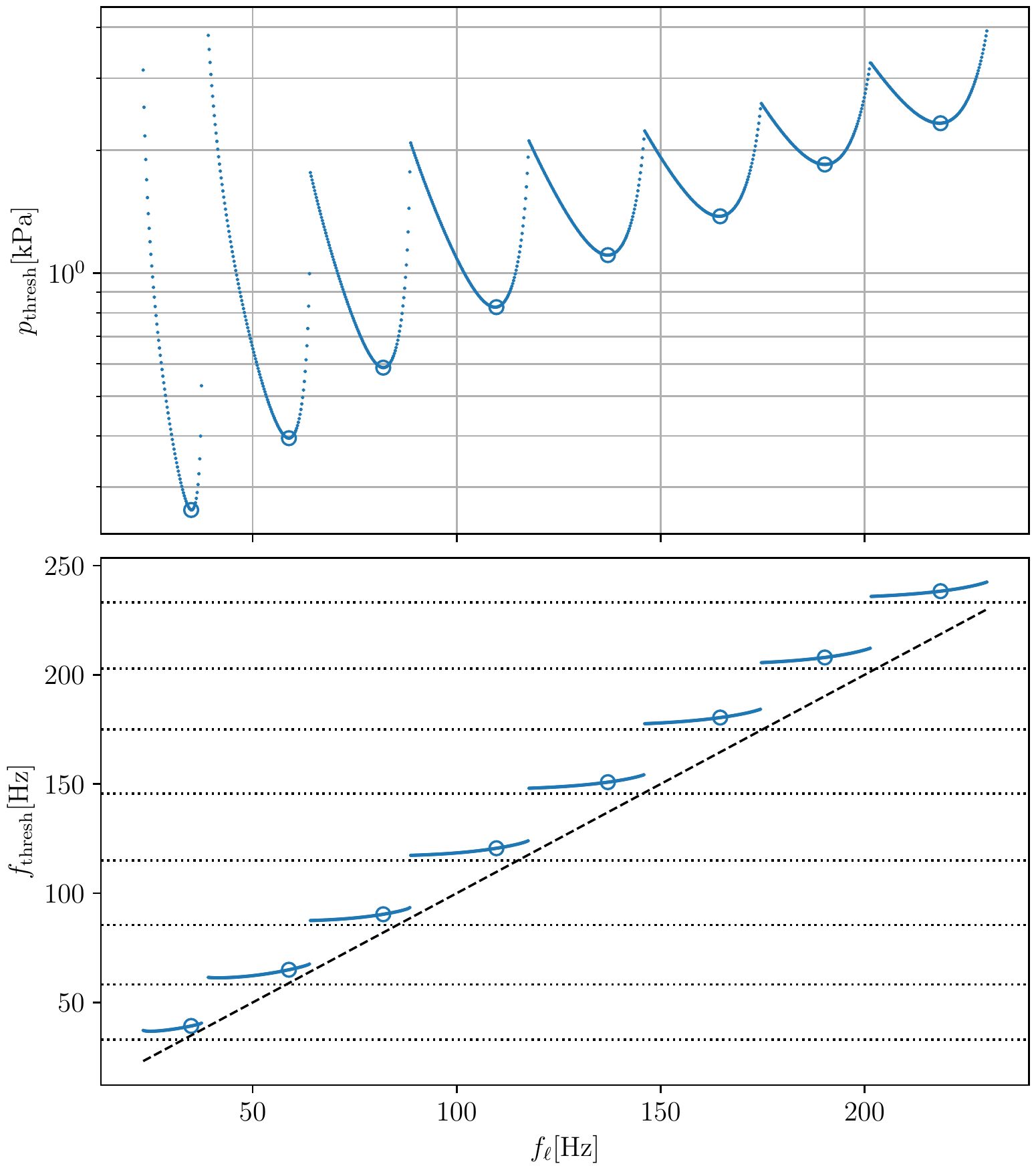}
    \caption{Results of the linear stability analysis applied to the B$\flat$-contrabass tuba. Top and bottom plots represent respectively the threshold mouth pressure and the threshold frequency with respect to the lips' resonance frequency. Circles point out the minima of each U-shaped pattern on the top plot. Horizontal dotted lines on the bottom plots locate the values of the acoustical resonances of the resonator; the black dashed line on the bottom plots represents $\fs{} = \fl$.}
    \label{asl_BE994}
\end{figure}

\begin{figure}[h!]
    \centering
    \includegraphics[width=0.7\linewidth]{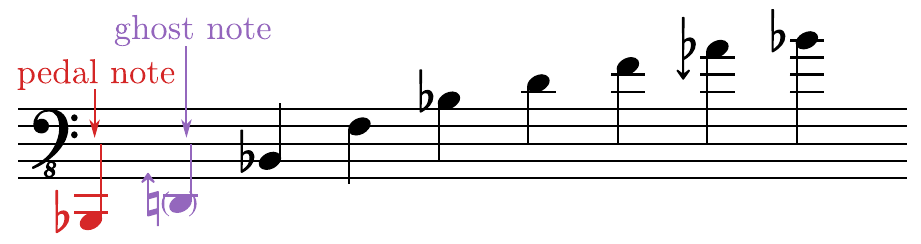}\includegraphics[angle=90,width=0.3\linewidth]{be994}
    \caption{Natural notes playable by a B$\flat$-contrabass tuba, that is to say without depressing any valves. From left to right: B$\flat$0 (pedal note), D1 (ghost note, naturally a bit sharp compared to an equal tempered scale), B$\flat$1, F2, B$\flat$2, D3, F3, A$\flat$3 (naturally a bit flat compared to an equal tempered scale), B$\flat$3.}
    \label{partiels_BE994_gn}
\end{figure}

The results of the linear stability analysis of the system are represented in figure \ref{asl_BE994} by the blue U-shaped patterns. As described in subsection \ref{NT}, these results give the values of the mouth pressure $p_m$ at which the equilibrium solution destabilises (the threshold mouth pressures), as well as the related frequency at which the oscillating solution emerges (the threshold frequencies). As a matter of fact, the multiple U-shaped patterns on the top plot of figure \ref{asl_BE994} reflect the fact that for a given configuration of the resonator (that is to say without depressing any valves), the tuba player is able to play several notes represented in figure \ref{partiels_BE994_gn} called \gu{natural notes}, just by changing the lips' resonance frequency (see for instance \cite{campbellScienceBrassInstruments2021}). In figure \ref{asl_BE994}, we chose to perform a stability analysis from D1 (37\,\text{Hz}, first U-shaped pattern) up to B$\flat$3 (233\,\text{Hz}, eighth U-shaped pattern) in terms of threshold frequency, even if a tuba player could play higher notes in practice. For each natural note or regime $n$ to emerge, there is an optimal lips' resonance frequency $\flo{eq}$ which corresponds to the minimum of a U-shaped pattern. This is associated with an optimal threshold mouth pressure $\pso{eq}$ and an optimal threshold frequency $\fso{eq}$. The notation $\flo{eq}$ has been introduced and used in \cite{matteoliMinimalBlowingPressure2021}, and is chosen to be as consistent as possible with \cite{velutHowWellCan2017}, in which the equilibrium optimal threshold value (namely the threshold given by the linear stability analysis) of a quantity $q$ in the $n^{\text{th}}$ regime was written $q_{\text{thresh},n}^{\text{opt}}$. Here, the threshold quantities are obtained either using linear stability analysis or through a bifurcation analysis, which will be defined and addressed in section \ref{BA}. Therefore, it has been chosen to add an extra superscript \gu{eq} for \gu{equilibrium}, referring to the linear stability analysis, or \gu{per} for \gu{periodic solutions}, referring to the analysis of bifurcation diagrams.

It is worth noting that the linear stability analysis is not able to reproduce the pedal note (red note in figure \ref{partiels_BE994_gn}), in contrast to the case of the trombone described in \cite{matteoliMinimalBlowingPressure2021}. Indeed, the first U-shaped pattern exhibits an optimal threshold frequency of 39\,\text{Hz} whereas the optimal threshold frequency of the second regime is 65\,\text{Hz}: there is a frequency interval of 8.8\,\text{semitones} between these two notes instead of the expected octave (12\,\text{semitones}) between a B$\flat$0 and a B$\flat$1. Moreover, it is inferred from recordings that the ghost note of the B$\flat$-contrabass tuba lies between a D1 and a D$\sharp$1, that is to say between a minor sixth (8\,\text{semitones}) and a perfect fifth (7\,\text{semitones}) under the second regime B$\flat$1. Thus, the first U-shaped pattern likely corresponds to the ghost note, which is thoroughly investigated in the following sections. Furthermore, it can be noticed that the patterns on the bottom plot of figure \ref{asl_BE994} are always above the line $\fs{} = \fl$. This is a characteristic of the outward-striking valve model: the instrument always plays a note slightly above the lips' resonance frequency.

\subsubsection{Bifurcation analysis}
\label{BA}

\begin{figure*}[h!]
    \centering
    \subcaptionbox{First Hopf bifurcation. \label{bd_BE994_reg1}}{\includegraphics[width=0.49\linewidth]{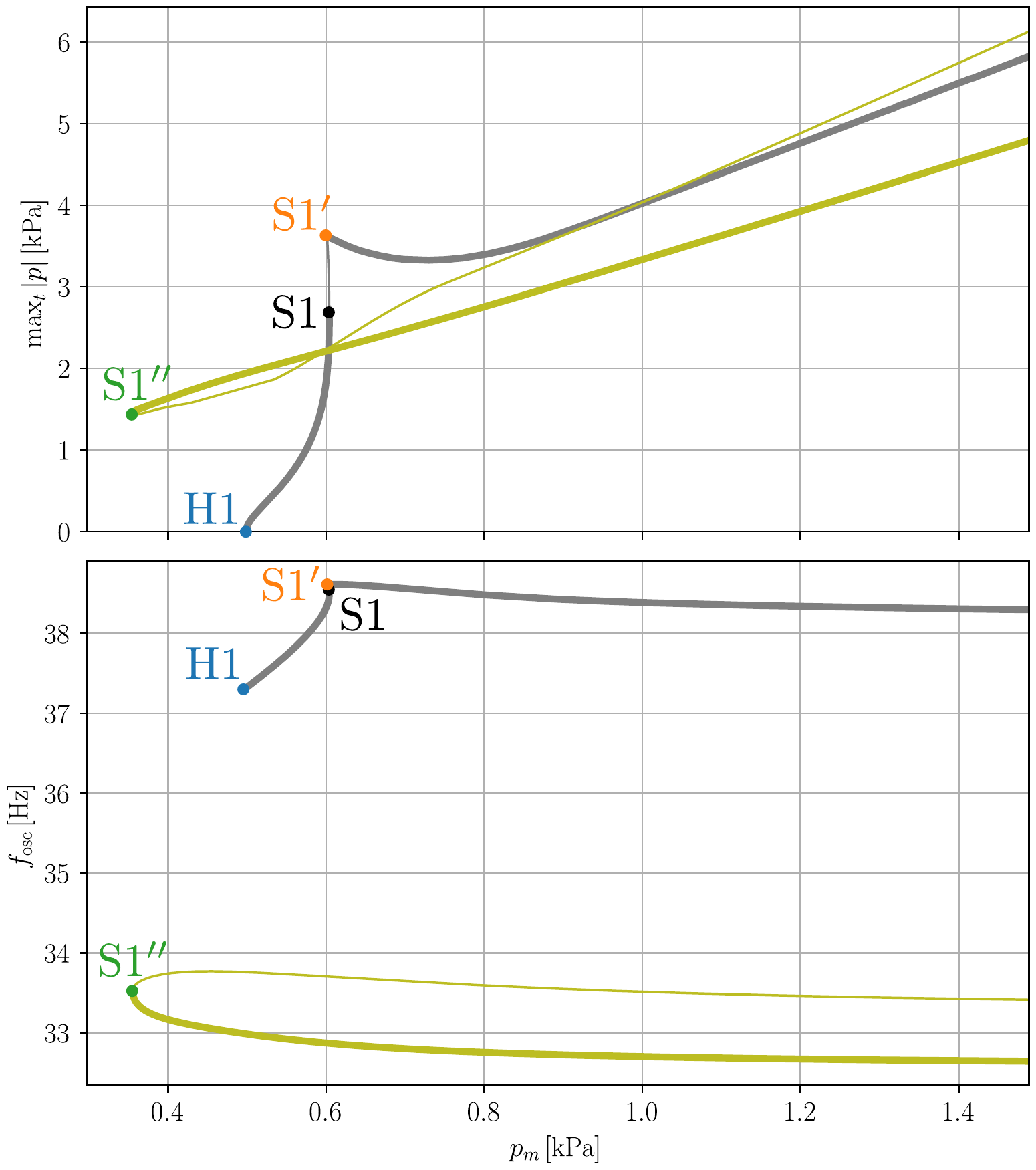}}
    \subcaptionbox{Second Hopf bifurcation. \label{bd_BE994_reg2}}{\includegraphics[width=0.49\linewidth]{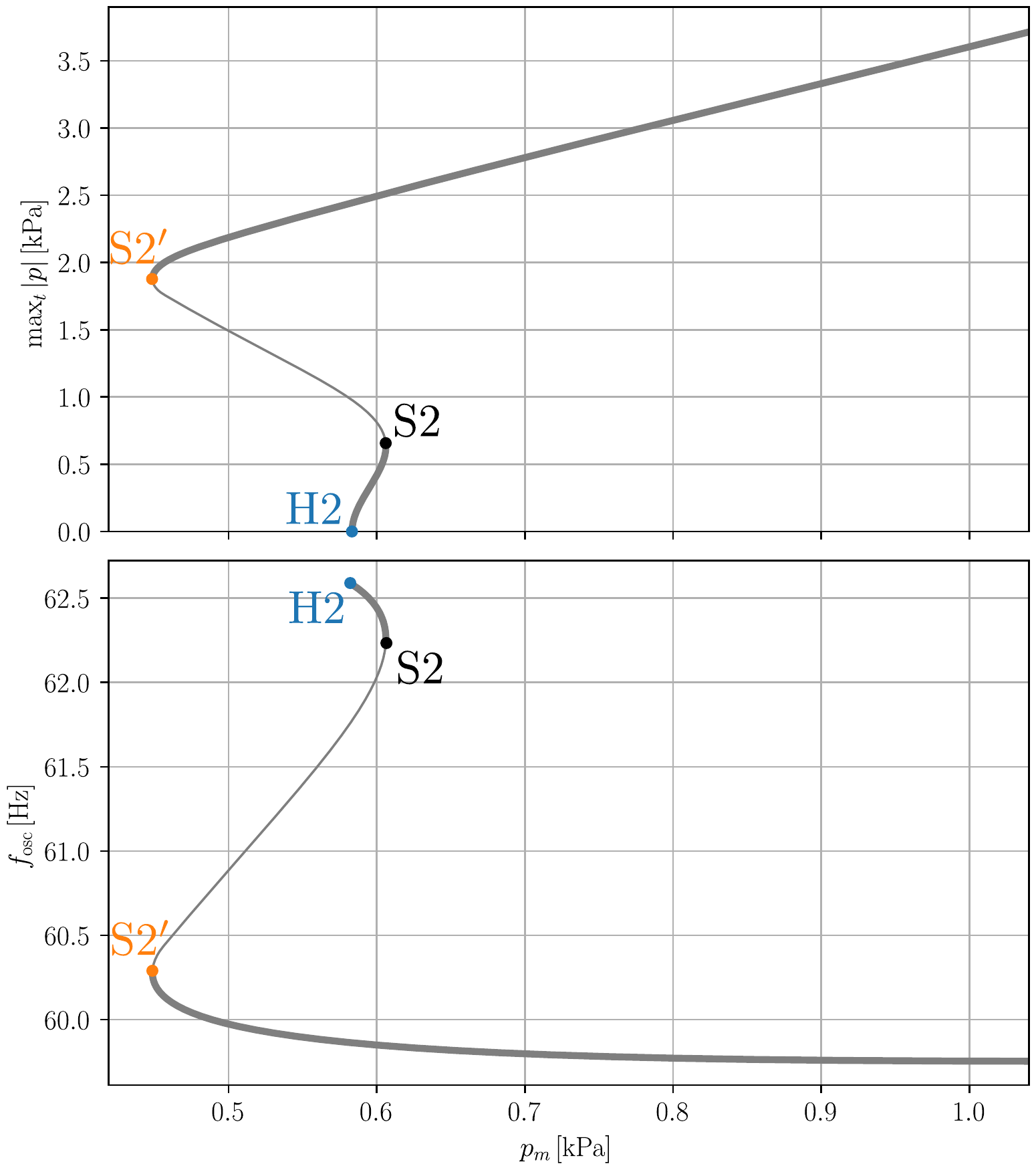}}
    \caption{Typical partial bifurcation diagrams for the first Hopf bifurcation and second Hopf bifurcation. Top and bottom plots represent respectively the maximum amplitude of the periodic oscillation branches and their oscillation frequency with respect to the blowing pressure. Left: case of the first Hopf bifurcation, $\fl = \flon{per}{\text{PN}} = 28.67\,\text{Hz}$; right: case of the second Hopf bifurcation, $\fl = 51.35\,\text{Hz}$. The line thickness indicates whether the branch portion is stable (thick line) or unstable (thin line). The points H1 and H2 correspond to the Hopf bifurcation points at which a stable oscillation regime arises from the equilibrium for the first and second Hopf bifurcation respectively. S1 and S2 correspond to the saddle-node bifurcation points at which the stable regime arising from the Hopf point destabilises. S1$'$ and S2$'$ correspond to the saddle-node bifurcation points at which another stable regime appears for the first and second Hopf bifurcation respectively. S1$''$ corresponds to the saddle-node bifurcation point at which the stable regime of the pedal note arises from the isolated branch (lime).}
    \label{bd_BE994_reg1_reg2}
\end{figure*}

In contrast to the linear stability analysis which gives information about the system close to its equilibrium only, the continuation method described in section \ref{NM} allows to investigate the system's behaviour arbitrarily far from it. In particular, it was shown in \cite{matteoliMinimalBlowingPressure2021} that the pedal note of an euphonium could actually be described thanks to this method. Figure \ref{bd_BE994_reg1_reg2} shows typical partial bifurcation diagrams representing the branch of oscillating solutions emerging from the first (left) and second (right) Hopf bifurcations of the B$\flat$-contrabass tuba.

Considering the second Hopf bifurcation (subfigure \ref{bd_BE994_reg2}) corresponding to a B$\flat$1 (58\,\text{Hz}), the upper bifurcation diagram exhibits a S-shaped branch emerging from the equilibrium point (or Hopf point) H1, which consists of two stable portions (thick lines) separated by an unstable portion (thin line). Such a curve means that the second Hopf bifurcation is actually related to the emergence of two stable regimes, and thus two notes. However, it is worth noting that the oscillation frequencies (bottom plot) taken at the emergence of each regime (namely the points H2 and S2$'$) are about 62.6\,\text{Hz} for H2 and 60.3\,\text{Hz} for S2$'$, less than a semitone away from H2. Therefore, both regimes are assumed to correspond to the same note, namely the B$\flat$1.

Regarding the first Hopf bifurcation (subfigure \ref{bd_BE994_reg1}), a part of the bifurcation diagram is discussed in the same way as the second Hopf bifurcation (namely the grey curve), since the oscillation frequencies taken at H1 and S1$'$ are less than a semitone away from each other\footnote{in contrast to the case of the trombone described in \cite{matteoliMinimalBlowingPressure2021}, in which the stable portion emerging from the Hopf point is almost non-existent given its narrow mouth pressure range of stability, as the stable portion emerging from the unstable portion actually corresponds to the pedal note.}. Both these regimes are considered to be only one, corresponding to the ghost note since the whole branch arises from the equilibrium. However, there also exists another stable regime represented by the isolated lime curve, whose oscillation frequency (taken at S1$''$) lies 2.4\,\text{semitones} away from the stable regime emerging from H1 and 1.8\,\text{semitones} away from the stable regime emerging from S1$'$. Given the significance of these intervals, the lime portion is considered to correspond to the pedal note -- namely B$\flat$0 -- of the B$\flat$-contrabass as pointed out in \cite{matteoliMinimalBlowingPressure2021}. It can be noted that this branch is not connected to the \gu{principal branch} (namely the branch arising from the Hopf point H1), making it more difficult to access without any prior work. As a matter of fact, the point S1$''$ is a saddle-node bifurcation point of the principal branch like S1$'$. However, it only connects to the principal branch above a critical value of the lips' quality factor $\Ql$, at which the two distinct branches on figure \ref{bd_BE994_reg1_reg2} merge into a unique branch. In the case of the B$\flat$-contrabass tuba for instance, this critical value lies around $\Ql = 21.7$. At such a value, S1$''$ is accessed in the exact same way as S1 or S1$'$, that is to say by performing a 1D-continuation (with respect to $p_m$) of the periodic solution branch from the Hopf point H1. Then, S1$''$ is continued a second time in the plane $\pr{\Ql,p_m}$ (2D-continuation) to obtain its locus at $\Ql = 7$ which is the value of interest in the present work.

In order to define the easiest notes playable according to the model and our definition of the ease of playing, the mouth pressure threshold of each solution listed above -- namely the mouth pressure at which a stable oscillating solution arises -- needs to be determined, as well as its corresponding threshold frequency. These minimal mouth pressures are given by the abscissa of the point S1$''$, H1 and S1$'$ on the top plot of subfigure \ref{bd_BE994_reg1} for the first Hopf bifurcation, and of H2 and S2$'$ on the top plot of subfigure \ref{bd_BE994_reg2} for the second Hopf bifurcation. However, it was demonstrated above that both for the first and second Hopf bifurcation, H$i$ and S$i'$ actually corresponded to the same note: the ghost note for the first Hopf bifurcation and B$\flat$1 for the second Hopf bifurcation. Taking the results in figure \ref{bd_BE994_reg1_reg2} as an example, the minimal mouth pressure of the ghost note is therefore given by H1, whereas the minimal pressure of B$\flat$1 is given by S2$'$. The points S1 and S2 have no relevance to our definition of the ease of playing, since they both correspond to a point at which an oscillating solution destabilises, meaning that they cannot be experimentally observed anymore. In that respect, they will not be taken into account in what follows.

\begin{figure}[h!]
    \centering
    \includegraphics[width=\linewidth]{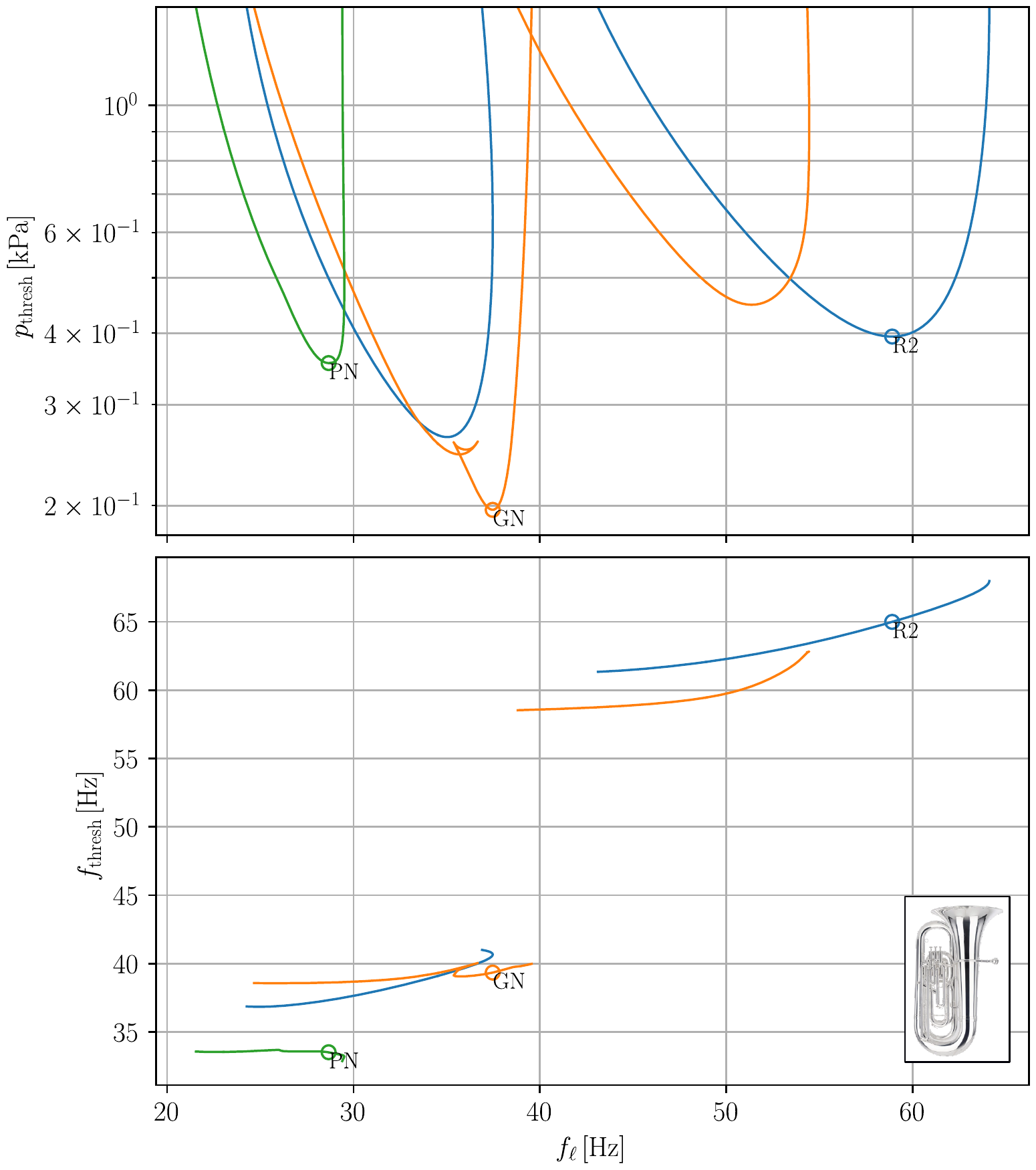}
    \caption{Top and bottom plots represent respectively the threshold pressures and threshold frequencies given by the locus of the Hopf points (blue) and the locus of the saddle-node bifurcation points (orange and green) with respect to the lips' resonance frequency in the case of the B$\flat$-contrabass tuba (colors correspond to the ones in figure \ref{bd_BE994_reg1_reg2}). The circles identify the local minimum of blowing pressure for each note, namely the pedal note (PN), the ghost note (GN), and the second regime (R2). In the case of the second Hopf bifurcation, the easiest note playable is given by the minimum of the Hopf points' locus (blue curve), since its minimal threshold mouth pressure is lower than the one given by the saddle-node bifurcation points' locus (orange curve). On the contrary, in the case of the ghost note it is given by the minimum of the saddle-node bifurcation points' locus (orange curve).}
    \label{seuils_BE994}
\end{figure}

The minimal mouth pressure of a note highly depends on the lips' resonance frequency, as we saw in figure \ref{asl_BE994} with the linear stability analysis. Therefore, the points S1$''$, H1, S1$'$, H2 and S2$'$ are continued in the plane $\pr{\fl,p_m}$, which is represented in figure \ref{seuils_BE994} in a similar manner as the results of the linear stability in figure \ref{asl_BE994}. In this figure, the blue U-shaped patterns are the locus in the plane $\pr{\fl,p_m}$ of the Hopf points H1 and H2 resulting from the continuation, and provide the same information as the two first U-shaped patterns in figure \ref{asl_BE994} obtained by performing a linear stability analysis. The orange U-shaped patterns correspond to the locus of S1$'$ and S2$'$\footnote{The angular loop that is seen on the locus of S1$'$ is called a swallowtail bifurcation \cite{seydelPracticalBifurcationStability2010}. Since hardly any physical interpretation can be given of this zone, we chose not to discuss it in this work.}, and the green one to the locus of S1$''$. For the second Hopf bifurcation, the easiest note playable -- namely the local minimum of the lowest U-shaped pattern -- is given by the locus of the Hopf points H2 (circle labelled \gu{R2}). On the contrary, it is given by the locus of the saddle-node bifurcation points S1$'$ in the case of the ghost note (circle labelled \gu{GN}) for the first Hopf bifurcation. There is no ambiguity for the easiest pedal note playable as there is only one U-shaped pattern (circle labelled \gu{PN}). The easiest pedal note, ghost note or second regime playable are then characterised by their individual optimal blowing pressure $\pso{per}$ (ordinate of the circled point on the top plot), its optimal lips' resonance frequency $\flo{per}$ (abscissa of the circled point on both plots), and its optimal threshold frequency $\fso{per}$ (ordinate of the circled point on the bottom plot) which corresponds to the playing frequency of the note. Since the U-shaped patterns cross paths at some points, the minimal mouth pressure $\ps{per}$ is given by the blue curve if it is below the orange one, and by the orange curve if it is below the blue one.

Apart from the case of the B$\flat$-baritone saxhorn which is studied in subsection \ref{BH}, the threshold pressures and threshold frequencies of the other tubas -- namely the C-contrabass tuba, the E$\flat$-bass tuba, the F-bass tuba, the B$\flat$-euphonium and the B$\flat$-bass saxhorn -- are all displayed in appendix \ref{MBPST} in the same manner as figure \ref{seuils_BE994}. Even though theses figures are interpreted in the exact same way as the figure \ref{seuils_BE994}, they display various different shapes of U-shaped patterns, due to presence of cusps and swallowtails \cite{seydelPracticalBifurcationStability2010}.

Eventually, it could be noted that the pedal note is supported by the first acoustic mode (that is to say its frequency is close to the pedal note's expected playing frequency), but the frequency of the ghost note does not correspond directly to one of the resonance frequencies. The latter observation on the ghost note appears similar to the case of the trombone's pedal note (playing frequency around 60\,\text{Hz}), sounding much higher than the first resonance frequency (around 38\,\text{Hz}, see the red curve in the bottom plot of figure \ref{comp_tb_sb_eu}). Another common characteristic of the tuba's ghost note and the trombone's pedal note is that, according to the model, both belong to a branch of periodic solutions emanating from the branch of equilibrium solutions (Hopf points). It is also worth pointing out that the pedal note does not fundamentally result from the cumulated contribution of higher acoustic modes, as it is commonly admitted. Indeed, \cite{velutNumericalSimulationProduction2014} proves that the pedal note of a trombone or a tuba does exist with a dummy input impedance containing only the first peak, based on time-domain simulations using the same brass model as in the present paper. However, this work also shows that the number of acoustic modes taken into account highly impacts the playing frequency of the pedal note: the more acoustic modes, the lower the playing frequency, bringing it closer to the expected playing frequency of the note.

\subsection{Case of the B$\flat$-baritone saxhorn}
\label{BH}

\begin{figure}[h!]
    \centering
    \includegraphics[width=\linewidth]{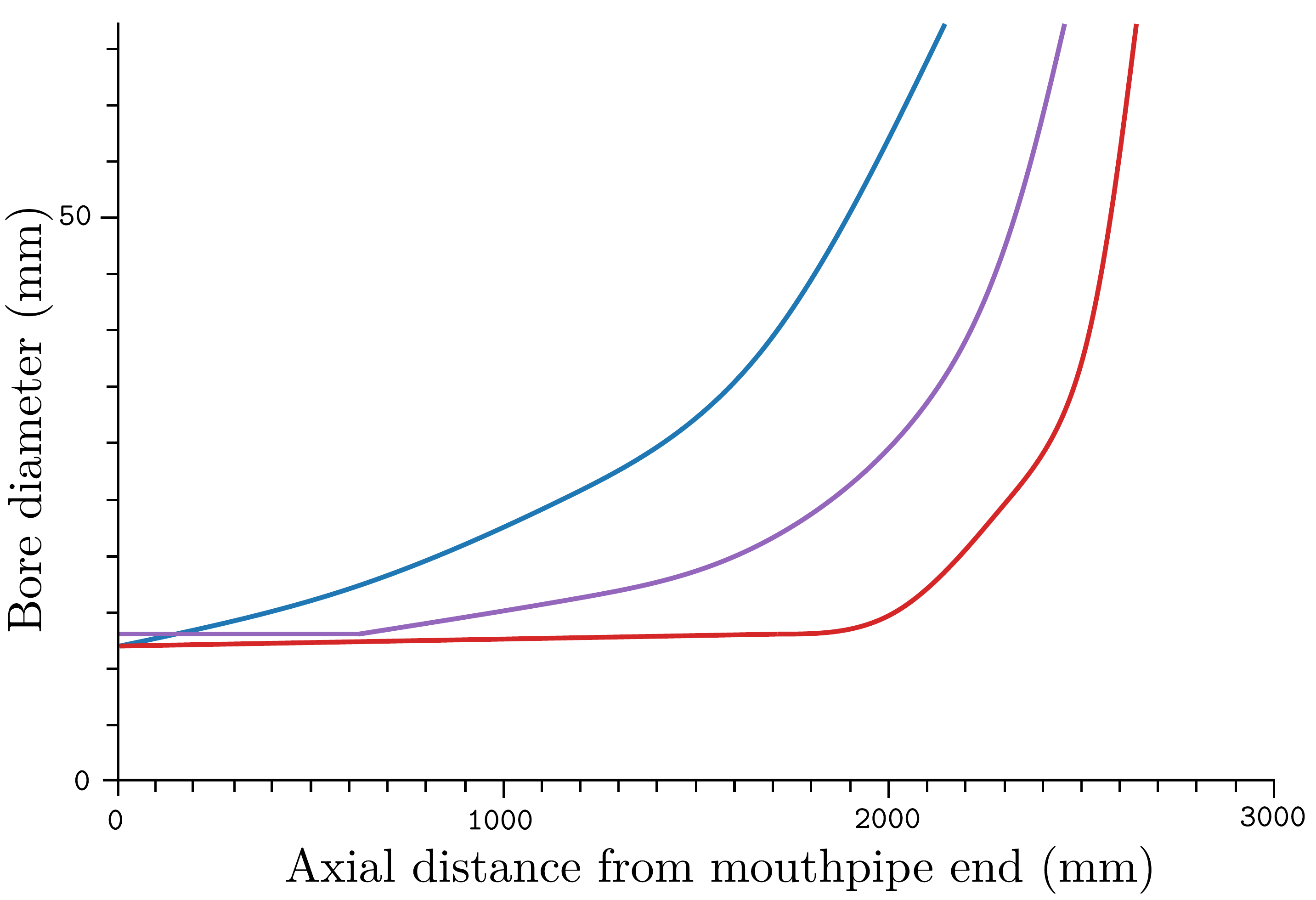}
    \vspace{2mm}\\
    \includegraphics[width=\linewidth]{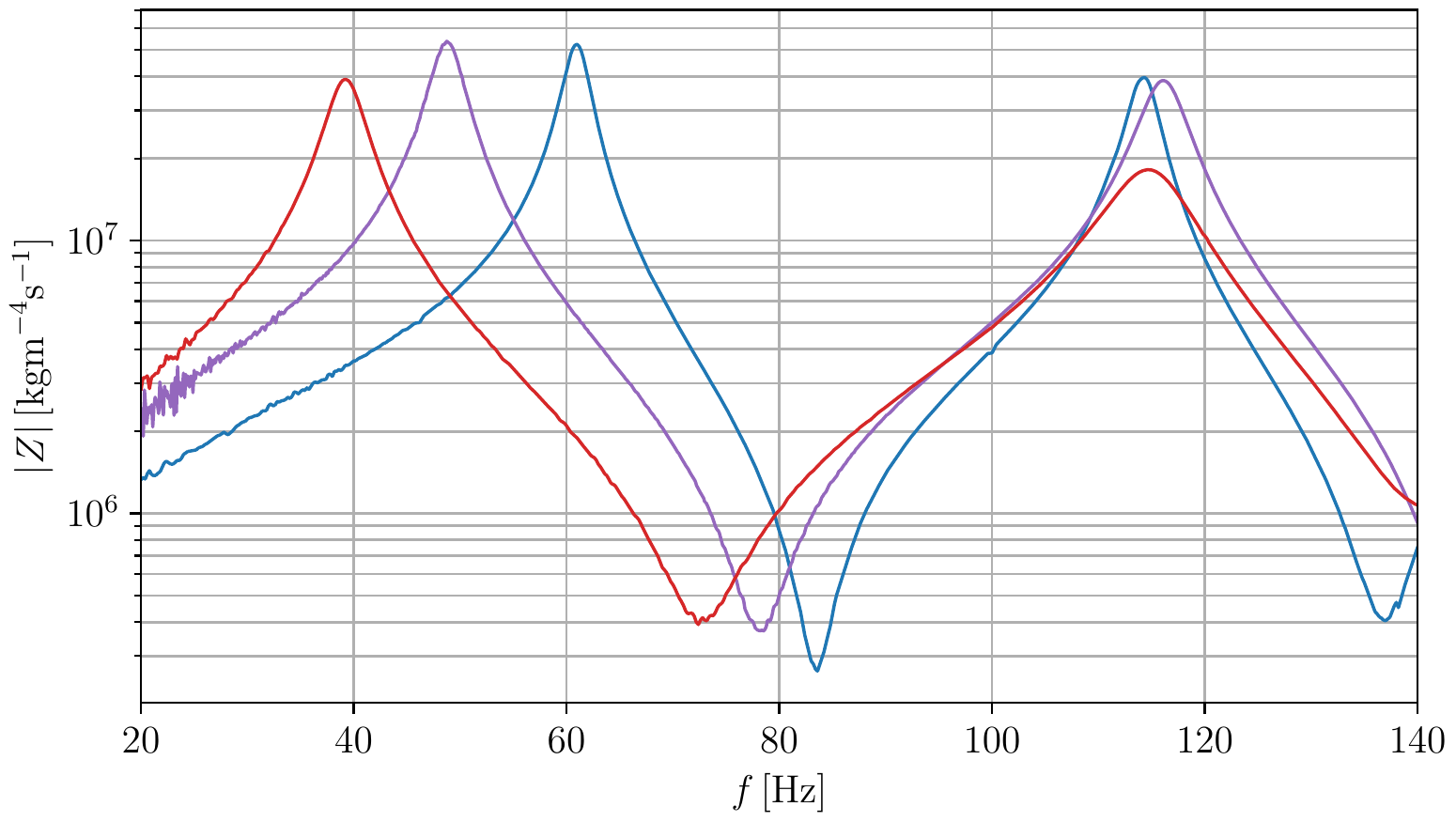}
    \caption{Top: bore profile as a function of the axial distance to mouthpipe end, taken from \cite{campbellScienceBrassInstruments2021} (where bore profiles of several other instruments can be found); bottom: modulus of the input impedance as a function of frequency (zoom on the first two peaks). Red: bass trombone; purple: B$\flat$-baritone saxhorn; blue: B$\flat$-euphonium.}
    \label{comp_tb_sb_eu}
\end{figure}

\begin{figure}[h!]
    \centering
    \includegraphics[width=0.7\linewidth]{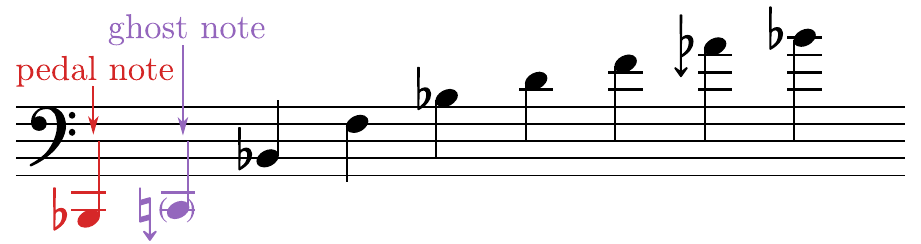}\includegraphics[angle=90,width=0.3\linewidth]{ba16}
    \caption{Natural notes playable by a B$\flat$-baritone saxhorn, that is to say without depressing any valves. From left to right: B$\flat$1 (pedal note), C2 (ghost note, perceived as a little flat), B$\flat$2, F3, B$\flat$3, D4, F4, A$\flat$4 (naturally a bit flat compared to an equal tempered scale), B$\flat$4.}
    \label{partiels_BA16_gn}
\end{figure}

This subsection focuses on the study of the first three natural notes of the B$\flat$-baritone saxhorn (B$\flat$1, C2, B$\flat$2, see figure \ref{partiels_BA16_gn}), which is one of the highest-pitched tubas among the seven tubas. Even though the baritone saxhorn still belongs to the family of bass brass instruments with predominantly-expanding bore profile, it is worth noticing that both its bore profile and the first peak of its input impedance lie between those of a tuba and a trombone, as it is highlighted in figure \ref{comp_tb_sb_eu}.

\begin{figure}[h!]
    \centering
    \includegraphics[width=\linewidth]{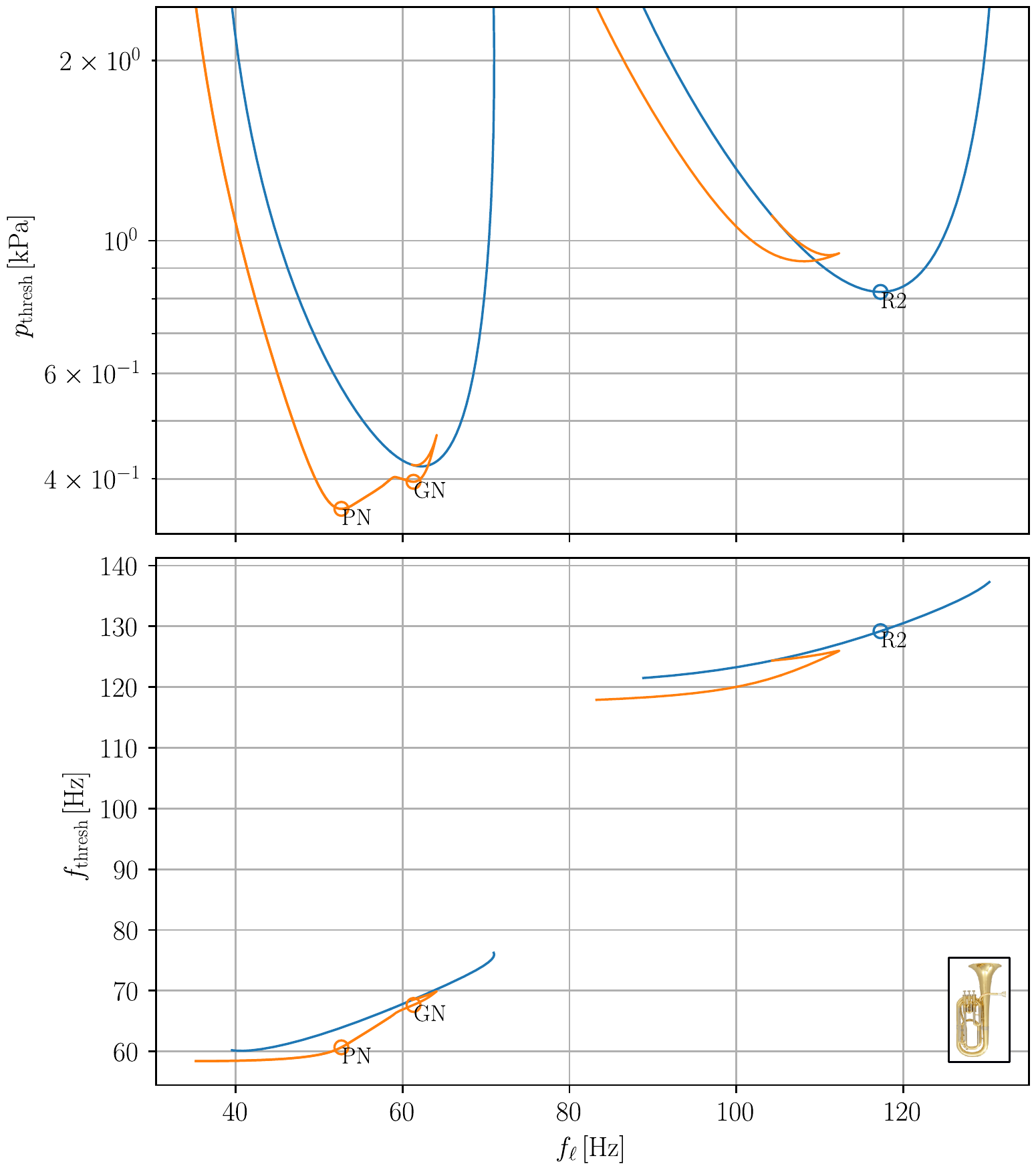}
    \caption{Top and bottom plots represent respectively the threshold pressures and threshold frequencies given by the locus of the Hopf points (blue) and the locus of the saddle-node bifurcation points (orange) with respect to the lips' resonance frequency in the case of the B$\flat$-baritone saxhorn. The circles identify the local minimum of blowing pressure for each note, namely the pedal note (PN), the ghost note (GN), and the second regime (R2). In the case of the second Hopf bifurcation, the easiest note playable is given by the minimum of the Hopf points' locus (blue curve), as its minimal threshold mouth pressure is lower than the one given by the saddle-node bifurcation points' locus (orange curve). In the case of the first Hopf bifurcation, the locus of the saddle-node bifurcation points (orange curve) shows two local minima, each one corresponding to the ghost note and the pedal note, as both these minima lie lower than the locus of the Hopf points and have a significantly different threshold frequency.}
    \label{seuils_BA16}
\end{figure}

In the case of the baritone saxhorn, the partial bifurcation diagram for the first Hopf bifurcation exhibits a S-shaped branch, similarly to the B$\flat$-contrabass tuba. Nevertheless, there seems to exist no isolated branch corresponding to the pedal note as in subfigure \ref{bd_BE994_reg1}. Instead, the S-shaped branch includes a stable portion emerging from a saddle-node bifurcation point S1$'$ corresponding to the pedal note, whereas the same portion of this branch in the case of the B$\flat$-contrabass tuba was only related to another specimen of the ghost note, as shown in subsection \ref{BA}. The locus of the saddle-node bifurcation points S$i'$ and Hopf points H$i$ in the plane $\pr{\fl,p_m}$ is displayed in figure \ref{seuils_BA16}. In the case of the second Hopf bifurcation, it can be noted that similarly to the B$\flat$-contrabass tuba, the easiest-to-play note is not given by the minimal mouth pressure of the saddle-node bifurcation points' locus, but by the minimal mouth pressure of the Hopf points' locus, as it lies lower. More importantly, the saddle-node bifurcation curve for the first Hopf bifurcation is W-shaped, namely it displays two local minima, each of them labelled PN and GN on the left side of figure \ref{seuils_BA16}. Their optimal threshold frequency on the bottom plot shows that they correspond to two different notes, as they differ by about a whole tone from each other. Furthermore, this W-shaped pattern having its minimal mouth pressures both lower than the minimal mouth pressure of the Hopf points' locus, they are considered to correspond to the pedal note and the ghost note, respectively. It is worth noting that, in contrast to the case of the B$\flat$-contrabass tuba studied in subsection \ref{BA}, for which the pedal note and the ghost note belong to a distinct U-shaped pattern, these notes belong to a unique W-shaped pattern in the case of the baritone saxhorn. Furthermore, in the case of the B$\flat$-contrabass tuba, there also exists a similar case of a W-shaped pattern for the first Hopf bifurcation (orange curve on the left side of figure \ref{seuils_BE994}) caused by a swallowtail, except here both local minima of this W-shaped pattern correspond to the ghost note, given their very close optimal oscillation frequency at threshold.

\section{Comparison between tubas with different nominal pitch}
\label{CBTWDTL}

In this section, the recording set-up used to experimentally assess the playing frequencies of the first three natural notes of a tuba is first presented. Second, the experimental playing frequencies of the seven tubas are compared to the optimal threshold frequencies obtained thanks to the numerical bifurcation analysis presented in section \ref{MBPPF}.

\subsection{Recording set-up}
\label{RS}

\begin{figure}[h!]
    \centering
    \subcaptionbox{Recording scheme. \label{rec_scheme}}{\includegraphics[width=0.49\linewidth]{recording_scheme_bmc}}
    \subcaptionbox{Exercise to play. \label{ex_ctrb}}{\includegraphics[width=0.49\linewidth]{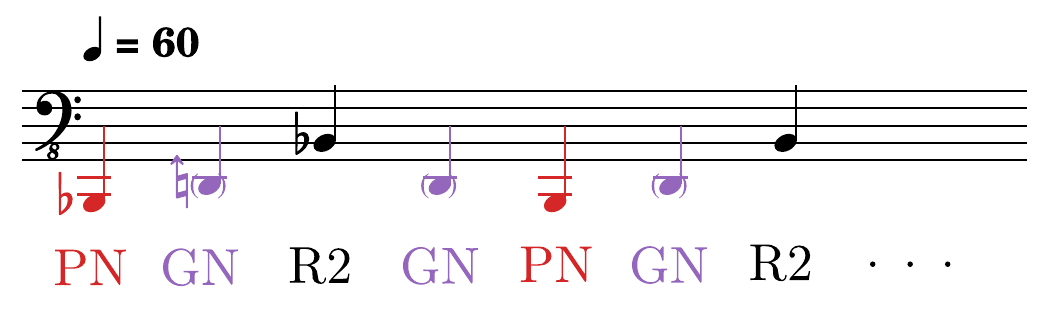}}
    \caption{Experimental scheme used to record the professional tuba players (left) and score of the exercise they were asked to play (right), here in the case of the B$\flat$-contrabass tuba. The audio recorder is a Zoom H4n, allowing stereo recording. The recording took place in a semi-anechoic room.}
    \label{experimental_scheme}
\end{figure}

\begin{figure}[h!]
    \centering
    \includegraphics[width=\linewidth]{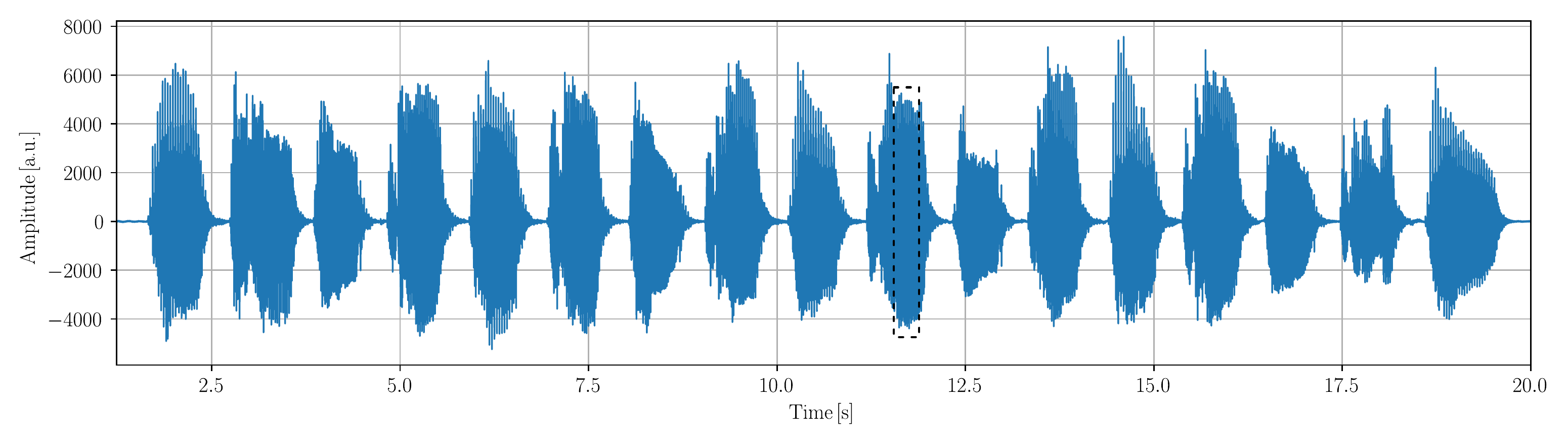}
    \includegraphics[width=\linewidth]{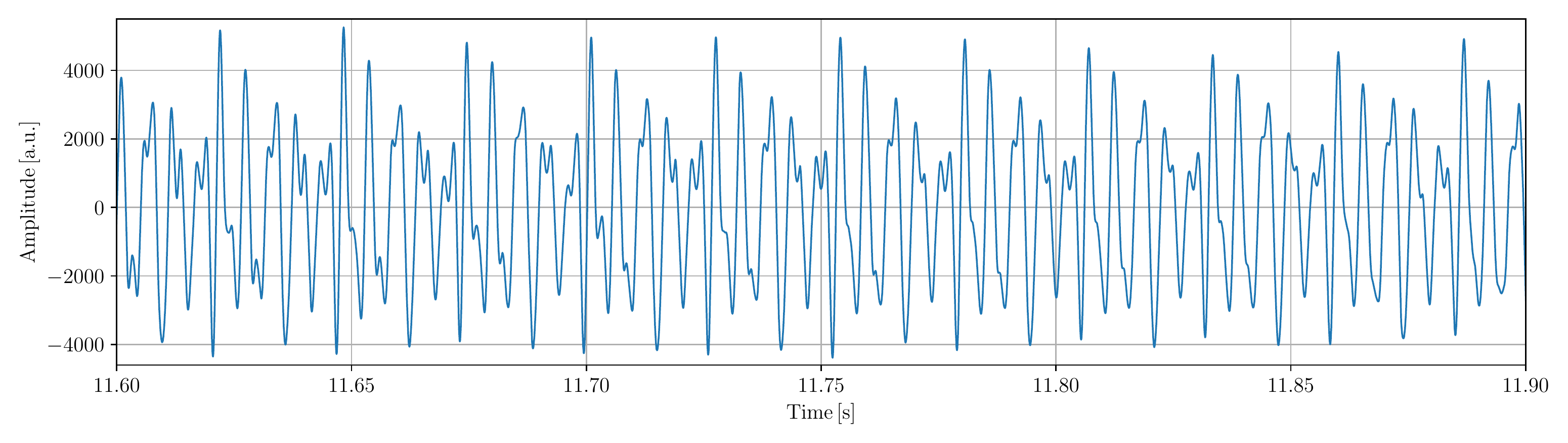}
    \caption{Top: shape of the time signal recorded from one tuba player playing the exercise described in figure \ref{experimental_scheme} on the B$\flat$-contrabass tuba: each \gu{wavepacket} corresponds to a note of the exercise. Bottom: details of the recorded signal of a ghost note played during the exercise, corresponding to the dashed box drawn in the top plot.}
    \label{Mathieu_BE994}
\end{figure}

The frequency of the easiest-to-play pedal note, ghost note and second regime was experimentally assessed based on recordings of two professional tuba players, whose experimental set-up is described in figure \ref{experimental_scheme}. They were both asked to play successively these three notes according to the procedure described in figure \ref{ex_ctrb} (all tuning slides completely pushed in). They were asked to play the easiest notes playable in terms of \gu{slotted notes} (which are defined here as the most stable notes) according to the player, regardless of the pitch. After recording the exercise using a Zoom recorder H4n in a semi-anechoic room\footnote{The signal displayed in figure \ref{Mathieu_BE994} can be listened to at \url{http://perso.univ-lemans.fr/~rmatte/bb-contrabass_tuba_exercise.mp4}.}, the instantaneous frequency of each note was extracted using Yin \cite{decheveigneYINFundamentalFrequency2002} based on the time signal of the note without its transient states, as displayed in figure \ref{Mathieu_BE994} (dashed box). In this way, the experimental conditions were as close as possible to the framework of the numerical bifurcation diagrams, which provide informations on permanent regimes only.

\subsection{Results}
\label{R}

The procedure described in section \ref{RS} was performed by both professional tuba players on the seven different tubas. The tubas were characterised in the following manner: from the lowest nominal pitch (B$\flat$-contrabass tuba) to the highest (baritone saxhorn), each tuba had first its input impedance measured thanks to the same device used in \cite{macalusoTrumpetNearperfectHarmonicity2011} -- which is seen in figure \ref{z_bridge} -- and was afterwards given to the professional tuba players to be played.

Several experimental factors such as temperature and hygrometry rate are known to have a significant impact on the pitch of a note, but were not taken into account in the numerical model. Moreover, the lips' parameters such as $\Ql$ were probably not held constant in the experiments. Therefore, comparing the notes' pitches given by the numerical bifurcation diagrams and the recordings would be irrelevant. Instead, we choose to compute the frequency interval in semitones between two notes, as the quantity to compare between the numerical results and the experimental results.

\begin{figure*}[h!]
    \subcaptionbox{Ghost note/second regime. \label{GN_R2}}{\includegraphics[width=0.49\linewidth]{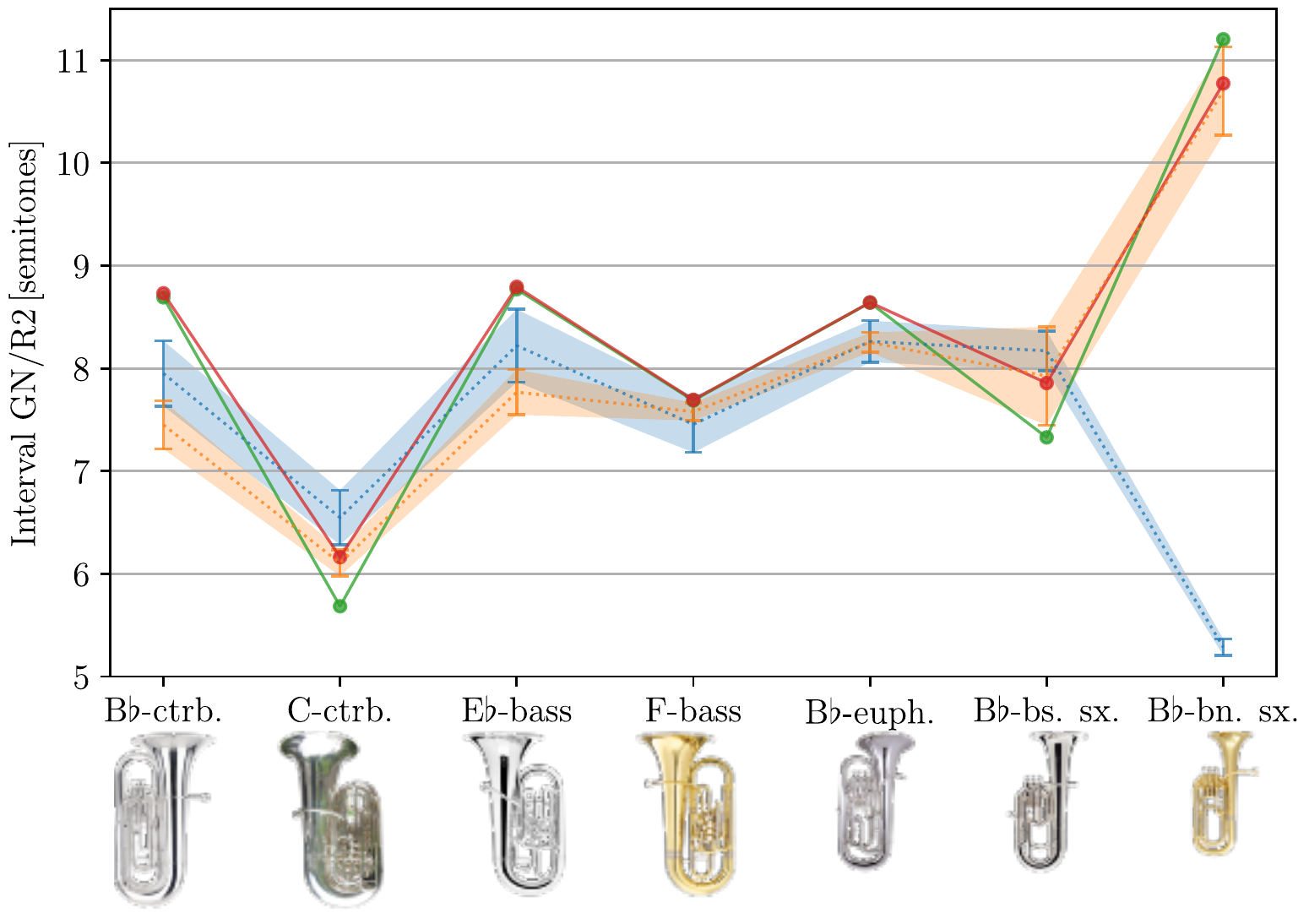}}
    \subcaptionbox{Pedal note/ghost note. \label{PN_GN}}{\includegraphics[width=0.49\linewidth]{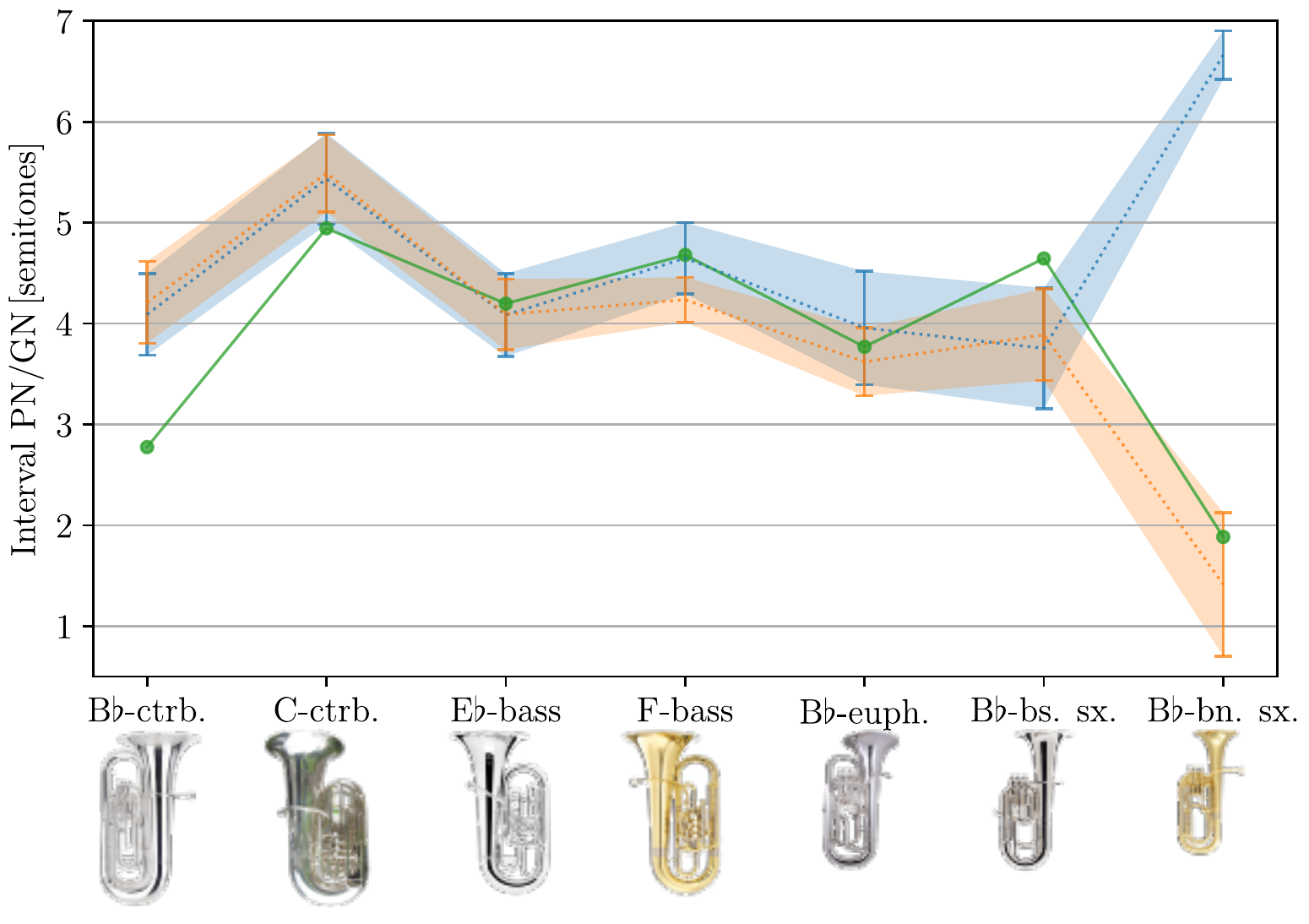}}\\
    \subcaptionbox{Pedal note/second regime. \label{PN_R2}}{\includegraphics[width=0.49\linewidth]{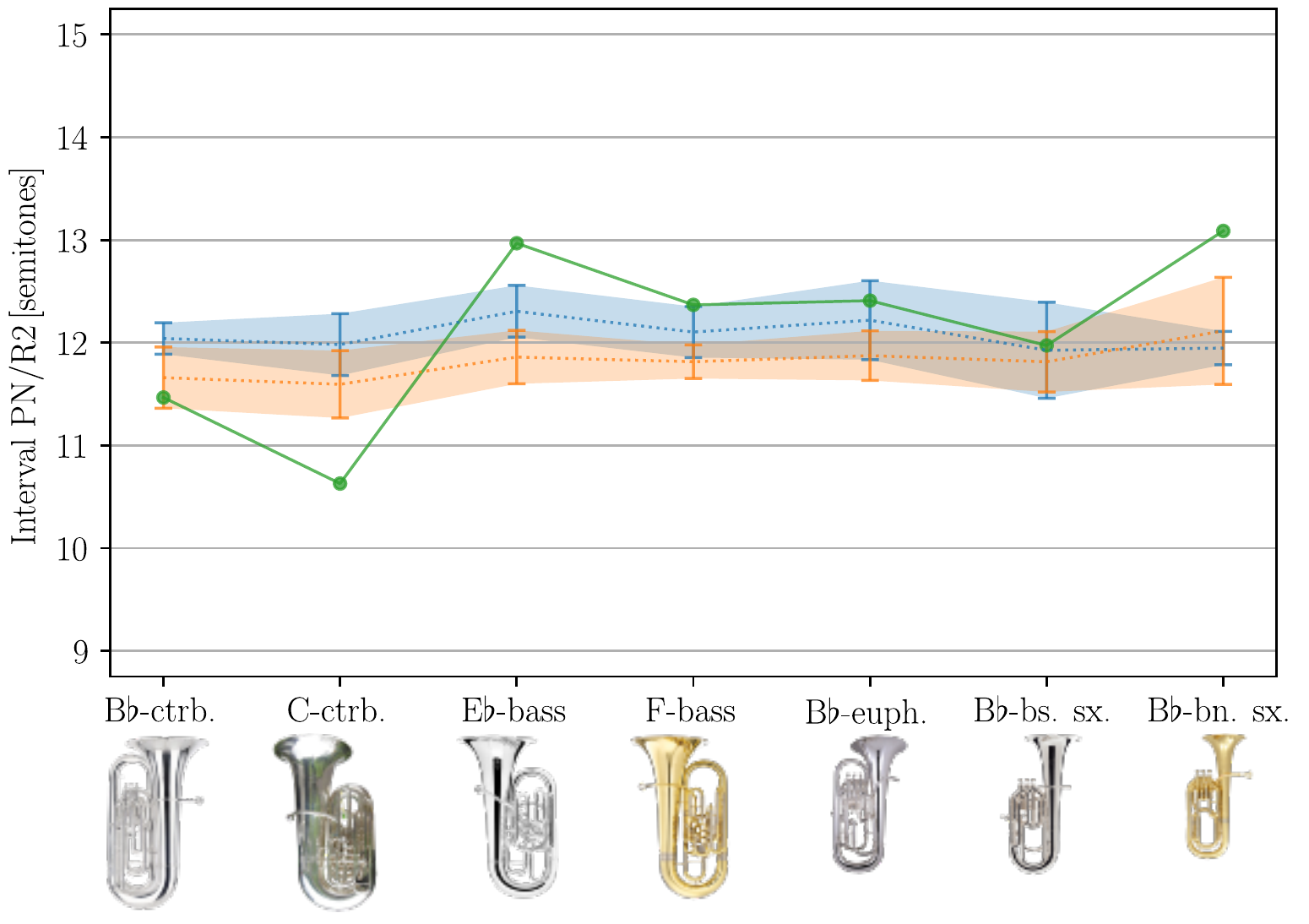}}
    \caption{Frequency intervals with respect to type of tuba. Blue points: frequency interval inferred from the recording of the first tuba player, orange points: frequency interval inferred from the recording of the second tuba player, green points: frequency interval inferred from the bifurcation diagrams, red points: frequency interval inferred from the linear stability analysis (when relevant). The instantaneous frequency (respectively the \gu{error} bar) of a note is computed taking the average (respectively the standard deviation) of the frequency over the notes occurrences in the exercise -- usually from 4 to 8 identical notes -- of the average frequencies individually computed by Yin \cite{decheveigneYINFundamentalFrequency2002} in a permanent regime.}
    \label{intervals}
\end{figure*}

Since three notes are involved, namely the pedal note PN, the ghost note GN and the second regime R2, three intervals are computed: $i_{\PN/\GN}$, $i_{\GN/\RD}$ and $i_{\PN/\RD} = i_{\PN/\GN}+i_{\GN/\RD}$. The results of these calculations are displayed in figure \ref{intervals}. A first interesting result is that, given the incursion of the experimental points, the monotonicity of the frequency intervals as a function of the tuba type is consistent between the recordings and the generic brass model. Secondly, the experimental frequency interval between the pedal note and the second regime (subfigure \ref{PN_R2}) tends to be always close to an octave (12 semitones). This could reflect the fact that, even told to chose the more slotted and easiest-to-play notes regardless of the pitch, the tuba players may have a tendency to instinctively play the pedal note in tune. On the contrary, the generic brass model exhibits a frequency interval between the pedal note and the second regime ranging from 10.6\,\text{semitones} in the case of the C-contrabass tuba to almost 13\,\text{semitones} in the case of the E$\flat$-bass tuba. Eventually, subfigure \ref{GN_R2} highlights the fact that the frequency interval computed using the linear stability analysis (red dots) is less than a quarter tone away from the frequency interval computed using the continuation method as a whole, meaning that the linear stability analysis is actually sufficient to describe the ghost note. As a matter of fact, the difference comes from the fact that the U-shaped pattern drawn by the locus of the saddle-node bifurcation points S$i'$ in the plane $\pr{\fl,p_m}$ has a lowest minimum than the one of the Hopf point, which is not predicted by the linear stability analysis. It is for instance the case for the ghost note of the B$\flat$ contrabass tuba, as shown in figure \ref{seuils_BE994}. The case of the B$\flat$-baritone saxhorn is singular, as the frequency intervals involving the ghost note (figure \ref{GN_R2} and \ref{PN_GN}) are not consistent between the two players. However, it is worth noticing that one of them seems to be consistent with the numerical model. This discrepancy could reflect the fact that a baritone saxhorn being close to a trombone -- for which the ghost note does not exist -- as regards its bore profile (see top plot of figure \ref{comp_tb_sb_eu}), the ghost note is not very well-defined. This is confirmed by the feeling, shared by both tuba players, that the ghost note on this instrument is not very slotted compared to the other tubas. Numerically, this could be illustrated by the fact that the ghost note and the pedal note both belongs to a W-shaped pattern whose minima are not very steep nor deep, making it easier to switch from the ghost note to the pedal note and vice-versa. In the case of the other tubas, each note was characterised by a distinct U-shaped pattern, thus making it more difficult to switch continually from the ghost note to the pedal note and vice-versa.

\section{Conclusion}
\label{C}

Most results in this paper highlight the usefulness of bifurcation diagrams analysis to understand the behaviours of the complete nonlinear model of brass instrument, both near and arbitrarily far from the oscillation threshold. Here, this in-depth analysis is applied to seven type of tubas. In contrast to the trombone, tubas have an extra natural note -- referred to as \gu{ghost note} -- between their pedal note (first playable note) and their second regime (one octave above). One special aspect of the ghost note is to be lying at a different interval with respect to the other natural notes, depending on the instrument's bore geometry. Nevertheless, not only does bifurcation analysis prove that the considered elementary model of brass instrument is able to reproduce the whole range of natural notes of a tuba, but also accurately describes the diversity of ghost notes among seven tubas featuring different bore geometries. Our results show that this is in excellent agreement with recordings of two professional tuba players. 

The study especially focuses on the first three natural notes of the tubas, namely the pedal note, the ghost note and the second regime. The ease of playing of a tuba is chosen to be assessed based on the minimal pressure thresholds, namely the minimal value of mouth pressure at which an oscillating solution exists. Because the pedal note requires to determine oscillating regimes far from the equilibrium state, a linear stability analysis is not sufficient to characterise this note. Instead, the minimal pressure thresholds are estimated thanks to a bifurcation analysis. At the same time, the associated threshold oscillation frequency is inferred, which is directly related to the pitch of the note. For the considered elementary model of brass instrument, this method provides information on all the periodic regimes accessible to a tuba player for a given fingering. From the experimental point of view, the playing frequency of a note is estimated based on the recording of two professional tuba players.

The intervals between the pedal note and the ghost note, the ghost note and the second regime, and the pedal note and the second regime appear consistent between recordings and numerical bifurcation analysis for six tubas out of seven. The last tuba, the baritone saxhorn (which is one of the highest-pitched), actually has an intermediate bore profile between a tuba and a trombone, the latter featuring no ghost note. In this case, the two professional tuba players felt that the ghost note was not as \gu{slotted} (meaning that the player does not have the sensation of a stable, well-defined pitch) as on the other tubas, resulting in a discrepancy between the two tuba players for this instrument. Nonetheless, this fact is consistent with the bifurcation diagrams analysis, which points out that the ghost note is more easily accessible by bending the pedal note up (that is to say by physically raising the lips' resonance frequency) than it is on the other tubas. The ghost note is thus numerically less well-defined on the baritone saxhorn, as confirmed by the sensation of both professional tuba players.

One important limitation of the present study lies in the difficulty to estimate the lips' parameters used in the model. Indeed, it would be reasonable to think that lips' parameters -- such as $\Ql$, which is known to have a strong impact on the shape of the bifurcation diagrams -- would vary from playing a low note to playing a high note, not to mention from playing a contrabass tuba to playing a baritone saxhorn. Yet, these lips' parameters being difficult to measure both on an artificial mouth or directly on a player, they were kept constant regardless the instrument or the note played. Even though the obtained results look reasonable, that is to say consistent with the musicians’ experience, \emph{in vivo} measurements of lips parameters during musical performance would be very valuable. Furthermore, the harsh modelling of the contact between the lips is known to artificially add harmonics to the time signals generated. This could be avoided by considering a more realistic contact model between the lips, even though regarding the frequency intervals, the model presented in this paper yields satisfying results as it is. Eventually, it is worth noting that the low number of tuba players (two) and the note occurrences (four to eight) per exercise performed by each tuba player may affect the quality of the statistics yielded. Thus, the recording process described in this paper would also highly benefit from being undertaken with a higher number of players in order to assess its repeatability and to refine the results obtained regarding the comparison of the frequency intervals between the elementary brass model and the recordings.

\section{Acknowledgments}

We wish to thank both professional tuba players Mathieu Chalange and Matthias Quilbault for allowing us to measure their tubas and record their performances, as well as Julia Mourier for useful discussions and for her help in this work.

\begin{appendix}

\section{Minimum blowing pressures of the seven tubas}
\label{MBPST}

In this appendix are displayed the same plots as in figure \ref{seuils_BE994} (namely the threshold pressure and threshold frequency as functions of the lips' resonance frequency) for all the tubas that have not been treated in detail in this paper.

\begin{figure*}[h!]
    \subcaptionbox{C-contrabass tuba. \label{seuils_HBS_510}}{\includegraphics[width=0.32\linewidth]{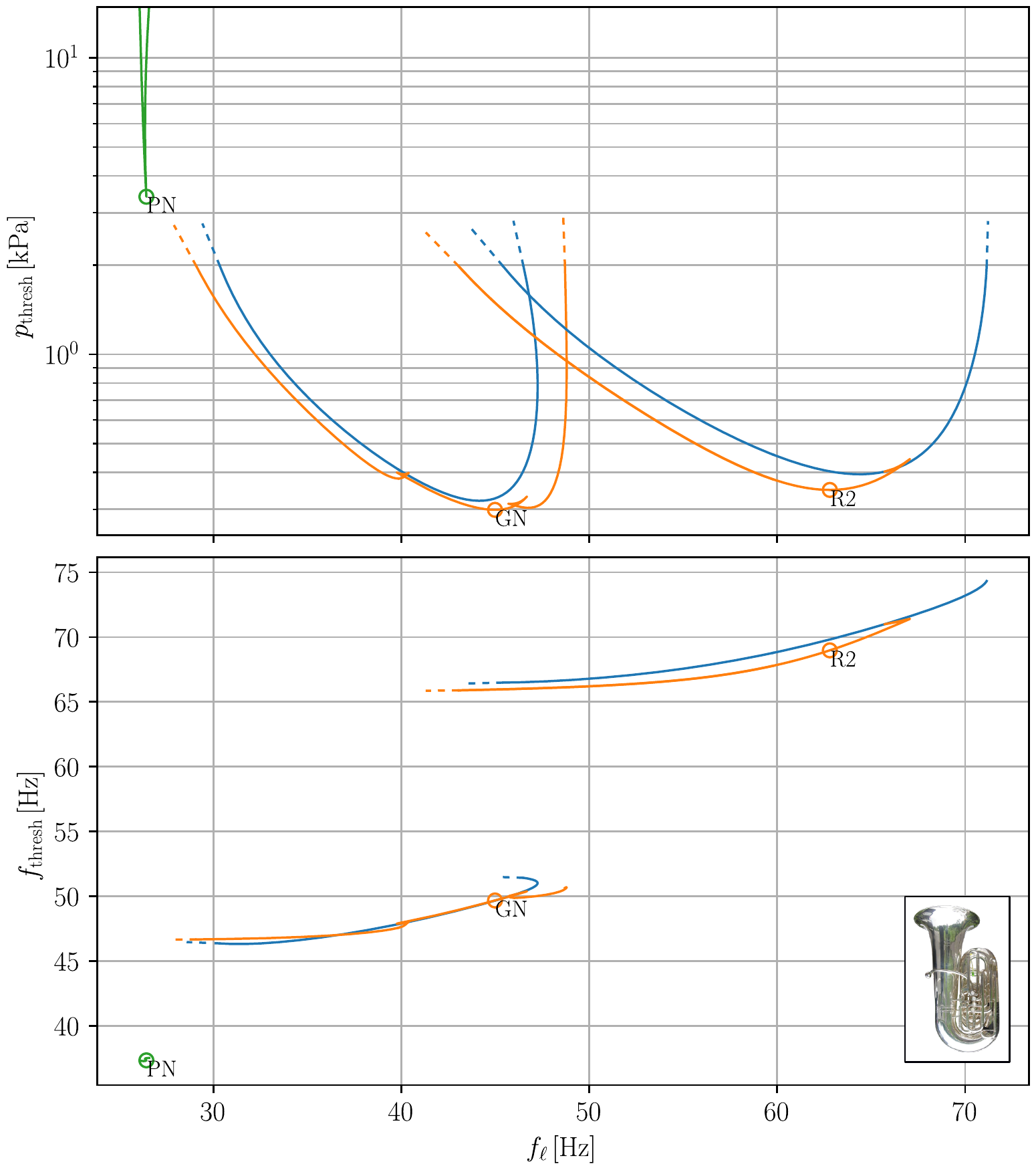}}
    \subcaptionbox{E$\flat$-bass tuba. \label{seuils_BE983}}{\includegraphics[width=0.32\linewidth]{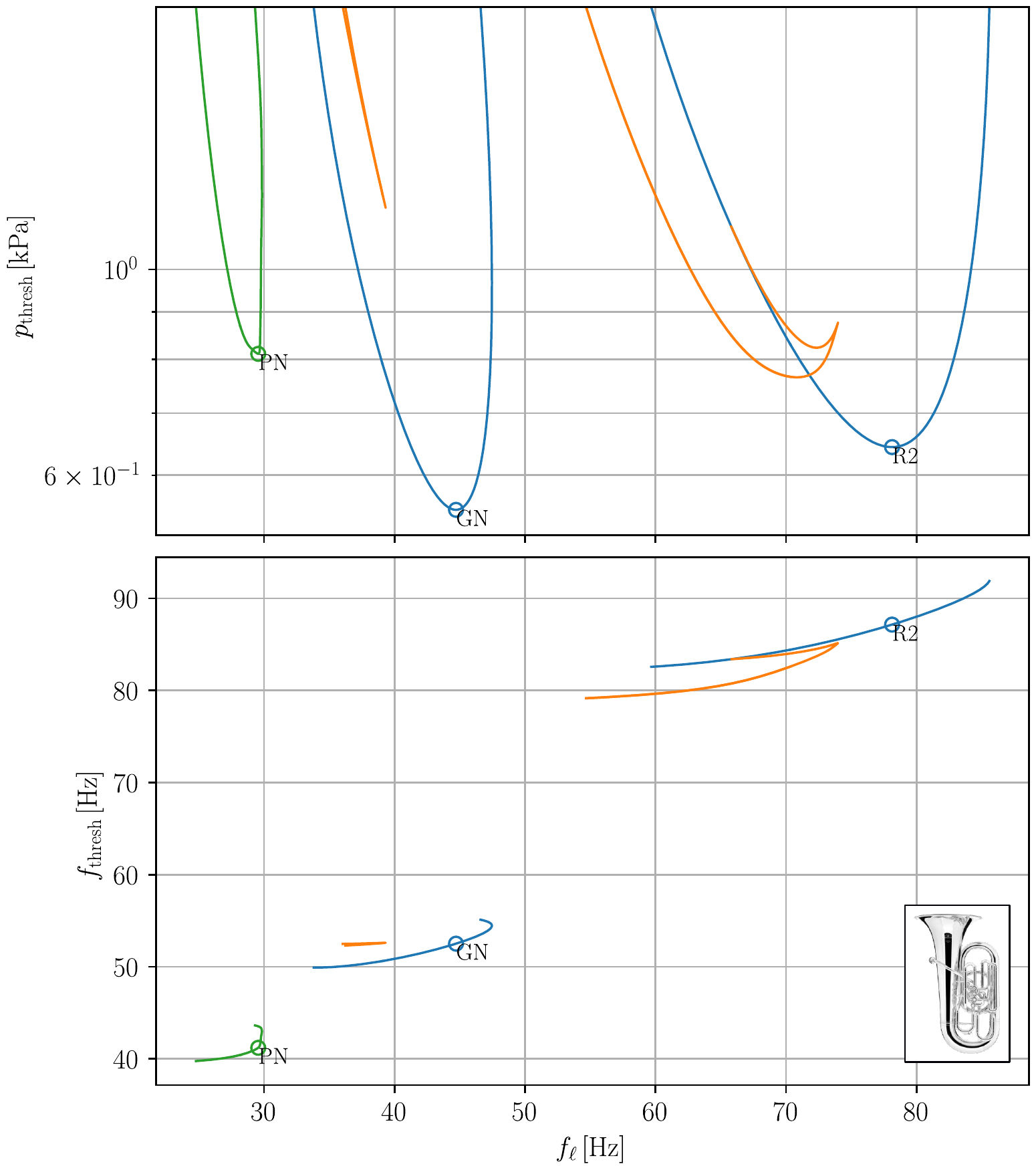}}
    \subcaptionbox{F-bass tuba. \label{seuils_MW_2250}}{\includegraphics[width=0.32\linewidth]{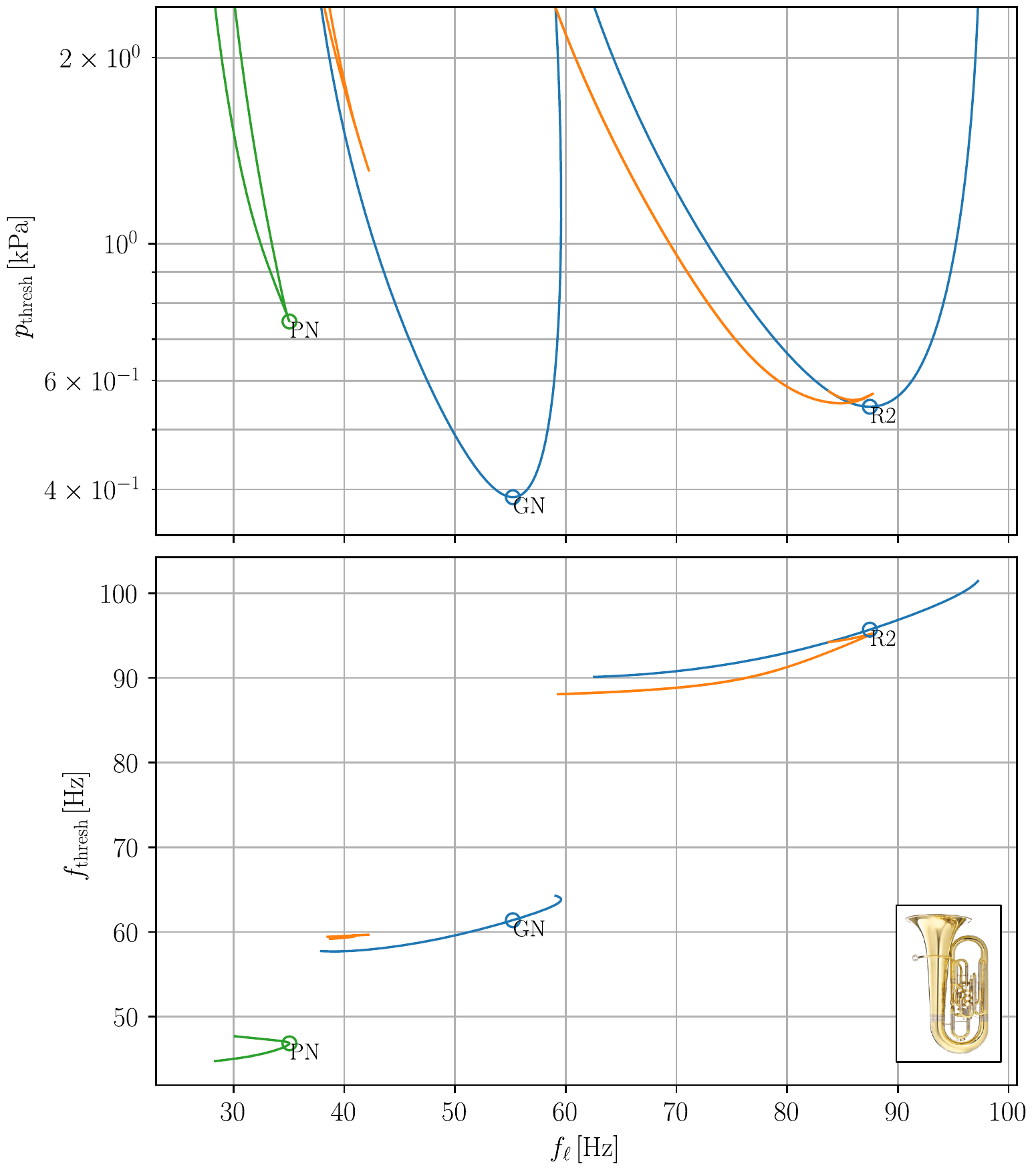}}\\
    \subcaptionbox{B$\flat$-euphonium. \label{seuils_BE967}}{\includegraphics[width=0.32\linewidth]{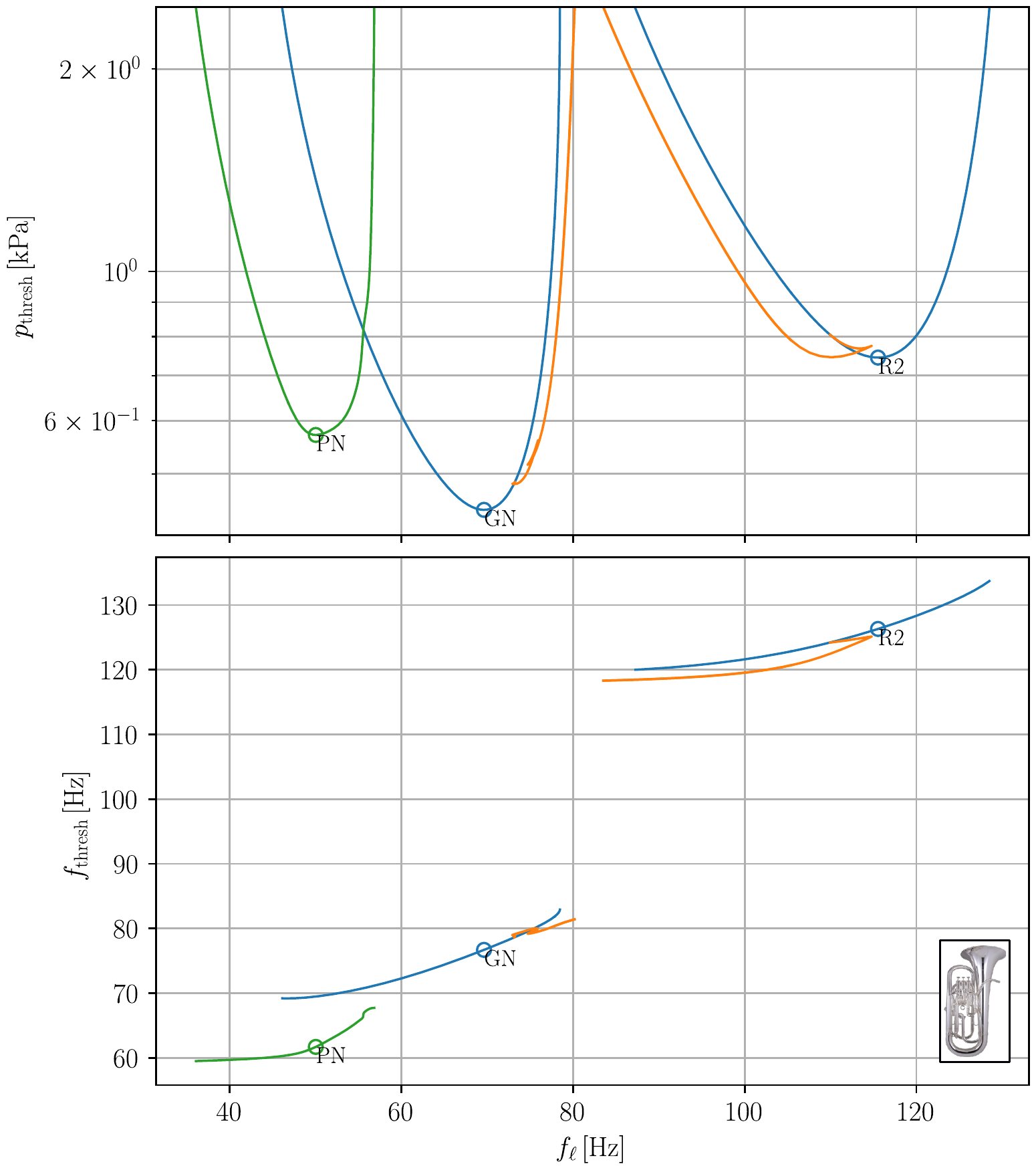}}
    \subcaptionbox{B$\flat$-bass saxhorn. \label{seuils_AC164}}{\includegraphics[width=0.32\linewidth]{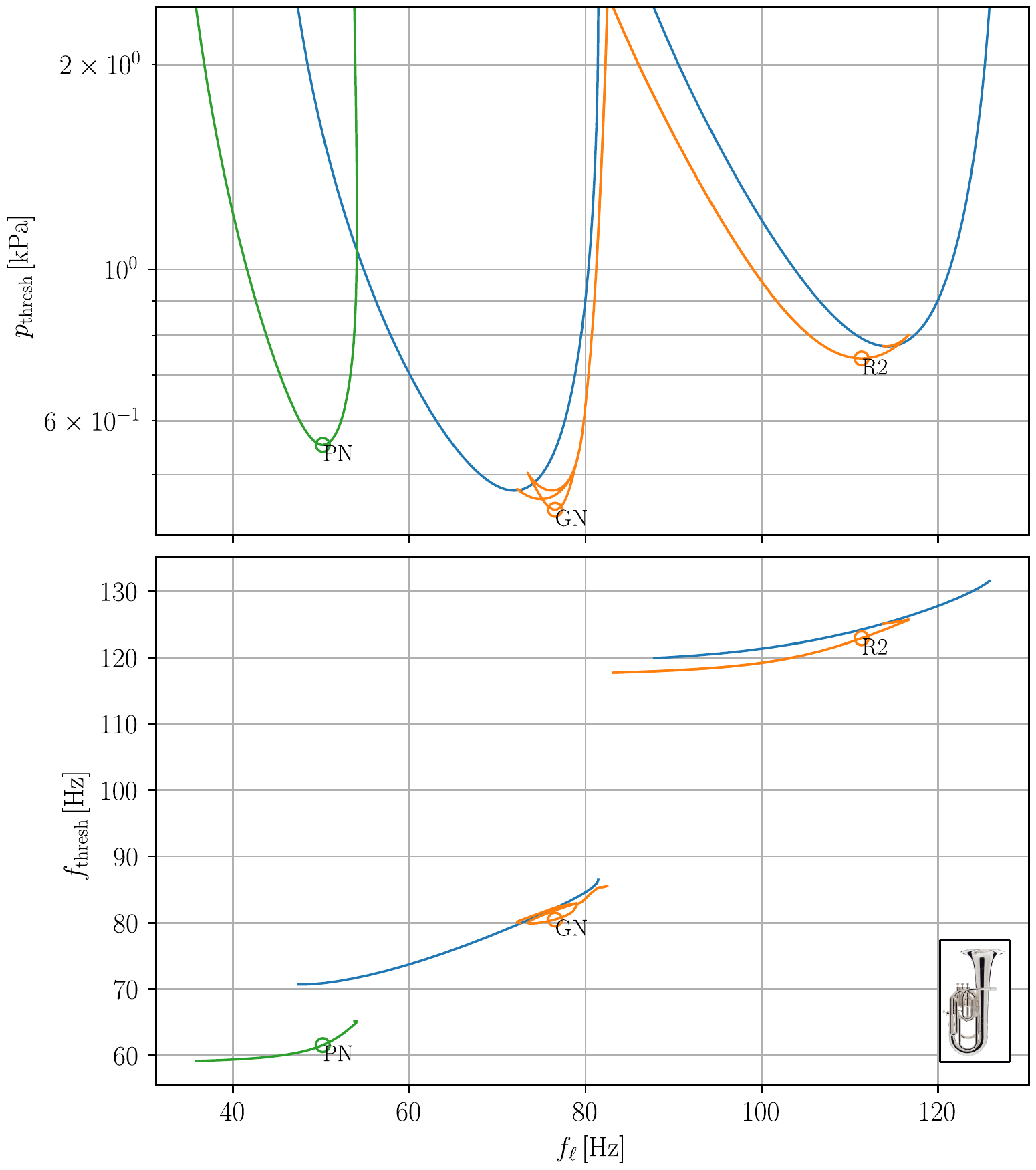}}
    \caption{Top and bottom plots represent respectively the threshold pressures and threshold frequencies given by the locus of the Hopf points (blue) and the locus of the saddle-node bifurcation points (orange and green) with respect to the lips' resonance frequency in the case of the C-contrabass tuba, E$\flat$-bass tuba, F-bass tuba, B$\flat$-euphonium and B$\flat$-bass saxhorn. The circles identify the local minimum of blowing pressure for each note, namely the pedal note (PN), the ghost note (GN), and the second regime (R2). The dashed portions at the ends of each U-shaped patterns in the case of the C-contrabass tuba mean that these curves still exist above $p_m = 2\,\text{kPa}$, but the computation was simply not performed above this value of mouth pressure.}
    \label{seuils_tubas}
\end{figure*}

These figures illustrate well the fact that for a given note, the easiest note playable correspond to the minimum of either the Hopf points' locus (blue curve), or the saddle-node bifurcation points' locus (orange curve). For instance, both the easiest ghost note and second regime playable correspond to the minima of the blue U-shaped patterns in the case of the B$\flat$-euphonium (subfigure \ref{seuils_BE967}), whereas they both correspond to the minima of the orange U-shaped patterns in the case of the B$\flat$-bass saxhorn (subfigure \ref{seuils_AC164}).

It is also worth noticing that, in the case of the C-contrabass tuba (subfigure \ref{seuils_HBS_510}), the pedal note's U-shaped pattern appears very narrow and have a very high optimal threshold mouth pressure (more than 3\,\text{kPa}) compared to the U-shaped patterns of the ghost note and the second regime. This would mean that the pedal note is very slotted, namely very stable and well-defined, but also hard to obtain considering the high mouth pressure the player would have to exert in the instrument, and the narrow span in the values of lips' resonance frequency at which this note exists. However, as discussed in subsection \ref{R} and section \ref{C}, several key-lips' parameters such as $\Ql$ were held constant regardless of the tuba in the bifurcation analysis, which is very unlikely to be the case in practice. Therefore, it could result in such a discrepancy.

Secondly, several regions exhibit a singular behaviour in figure \ref{seuils_tubas}. For instance, F-tuba's and E$\flat$-bass tuba's orange U-shaped patterns (subfigures \ref{seuils_MW_2250} and \ref{seuils_BE983}) show what are referred to as \gu{cusps} in \cite{seydelPracticalBifurcationStability2010} on their right side. Also, several swallowtails \cite{seydelPracticalBifurcationStability2010} are seen, a noticeable one being on the left orange U-shaped pattern in the case of the B$\flat$-bass saxhorn for instance (subfigure \ref{seuils_AC164}). Since no physical interpretation of these areas could be made, we chose not to discuss them in this study.

\end{appendix}

\bibliographystyle{apalike}
\bibliography{tubas}

\end{document}